# Absence of heat flow in $\nu = 0$ quantum Hall ferromagnet in bilayer graphene

Ravi Kumar[1,*], Saurabh Kumar Srivastav[1,*], Ujjal Roy[1,*], Ujjawal Singhal[1], K. Watanabe[2], T. Taniguchi[2], Vibhor Singh[1], P. Roulleau[3] and Anindya Das[1†]

[1]*Department of Physics, Indian Institute of Science, Bangalore, 560012, India.*
[2]*National Institute of Material Science, 1-1 Namiki, Tsukuba 305-0044, Japan.*
[3]*SPEC, CEA, CNRS, Université Paris-Saclay, CEA Saclay, 91191 Gif sur Yvette Cedex, France.*

**The charge neutrality point of bilayer graphene, denoted as $\nu = 0$ state, manifests competing phases marked by spontaneously broken isospin (spin/valley/layer) symmetries under external magnetic and electric fields. However, due to their electrically insulating nature, identifying these phases through electrical conductance measurements remains challenging. A recent theoretical proposal introduces a novel approach, employing thermal transport measurements to detect these competing phases. Here, we experimentally explore the bulk thermal transport of the $\nu = 0$ state in bilayer graphene to investigate its ground states and collective excitations associated with isospin. While the theory anticipates a finite thermal conductance in the $\nu = 0$ state, our findings unveil an absence of detectable thermal conductance. Through variations in the external electric field and temperature-dependent measurements, our results suggest towards gapped collective excitations at $\nu = 0$ state. Our findings underscore the necessity for further investigations into the nature of $\nu = 0$.**

**Introduction.** The $\nu = 0$ state of the bilayer graphene (BLG) quantum Hall (QH) offers a fertile playground for realizing the symmetry-broken many-body ground states arising from the interplay between the isospin anisotropy of electron-electron and electron-phonon interaction, Zeeman energy, and electric field across the layers[1–13]. The electric field tunability of the isospin 'layer' degree of freedom offers an extra knob to explore the rich phase diagram of the $\nu = 0$ state in BLG[14–20]. Theoretically, the $\nu = 0$ state is believed to harbour the four different ground states[1] with different spin and valley (equivalent to layer or sublattice) degrees of freedom, which are ferromagnetic (F), canted antiferromagnetic (CAF), partially layer polarized (PLP), and fully layered polarized (FLP). In the ferromagnetic and canted antiferromagnetic phase, each electron of the unit cell occupies a different sublattice with the same and canted spin polarization, respectively[1]. In the partially layered polarized, both electrons of the unit cell are shared with both the sublattices with opposite spin, and in fully layered polarized, both electrons occupy the same sublattice (equivalently same layer) with opposite spin polarization. The spin polarization of these phases on different sublattices is schematically shown in Fig. 1a. While all four phases are electrically insulating in bulk, the ferromagnetic phase is expected to harbour the conducting helical edge modes at the physical boundary of the sample[21]. Similar competing phases are also expected at $\nu = 0$ state of monolayer graphene (MLG)[2]. Although several experimental observations based on the charge transport measurement support some of the proposed insulating phases in both MLG[21–29] and BLG[14–20], these measurements do not directly capture the spontaneous symmetry breaking of the spin or isospin degrees of freedom.

*These authors contributed equally: Ravi Kumar, Saurabh Kumar Srivastav, Ujjal Roy
†anindya@iisc.ac.in

In contrast to the charge transport measurement, thermal transport experiments are believed to be more sensitive in identifying the spontaneous symmetry breaking of spin or isospin degrees of freedom of BLG. For example, the heat flow is allowed via gapless collective excitations (spin or isospin) like Goldstone modes for the spontaneously continuous symmetry-breaking phases, whereas the heat flow will be blocked for a gapped spectrum of collective excitations. This is indeed proposed by Falko Pientka et al.[30], who claim that at low temperatures, heat can flow via gapless collective modes for the CAF and PLP, but it should vanish for the FLP and F phases, where the excitation spectrum is gapped. A similar proposal was also made for MLG[7,31]. Despite these theoretical predictions, a study of thermal conductance for the $\nu = 0$ state was lacking due to the technical challenges involved in thermal transport measurement at low temperatures. We overcome these challenges by successfully developing a thermal conductance measurement technique in the QH regime[32–34] that employs Jhonson-Nyquist noise to determine the temperature of the heat reservoirs at low temperatures. This technique was used to measure the electronic contribution of heat flow due to chiral edge channels in integer and fractional quantum Hall phases in graphene and bilayer graphene[35–39].

In the spirit of the theoretical proposal of Falko Pientka et al.[30], here we report the study of heat flow in the $\nu = 0$ state in dual-gated hBN encapsulated bilayer graphene devices. To measure the thermal conductance, we have utilized a device geometry where a metallic floating contact reservoir is connected to two independent QH regions, whose filling is controlled by a global graphite back gate[35–37]. To create a $\nu = 0$ region in this device structure, we employed an additional local gate such that we could tune our device with and without the $\nu = 0$ state. Further, at $\nu = 0$ state, the displacement field ($D$) was varied from the value of zero to $\sim 0.1$ V/nm with expected ground states of CAF and FLP, respectively[30]. The thermal conductance was measured from very low temperature, $20$ mK to $1$ K. At low temperature, in the absence of $\nu = 0$ state, the measured thermal conductance matches with the expected electronic contribution. Next, the device was set to $\nu = 0$ state using the local gate, and surprisingly, no additional detectable contribution to the thermal conductance was measured, and the results remained the same irrespective of the external electric field ($D$) as well as temperature. This confirms the absence of heat flow in the $\nu = 0$ state from the anticipated collective excitations of isospin.

**Device and experimental principle.** The device schematic and measurement setup are shown in Fig. 1b and 1c. The device consists of a hBN encapsulated BLG placed on two independent graphite back gates, a global graphite back gate - BG1, and a local graphite back gate - BG2, control the charge carrier density or the filling factors of the blue-colored and green-colored part in Fig. 1b and 1c, respectively. Details of the device structures and optical images are shown in the supplementary information (SI)-S1 and SI-Fig. 1.

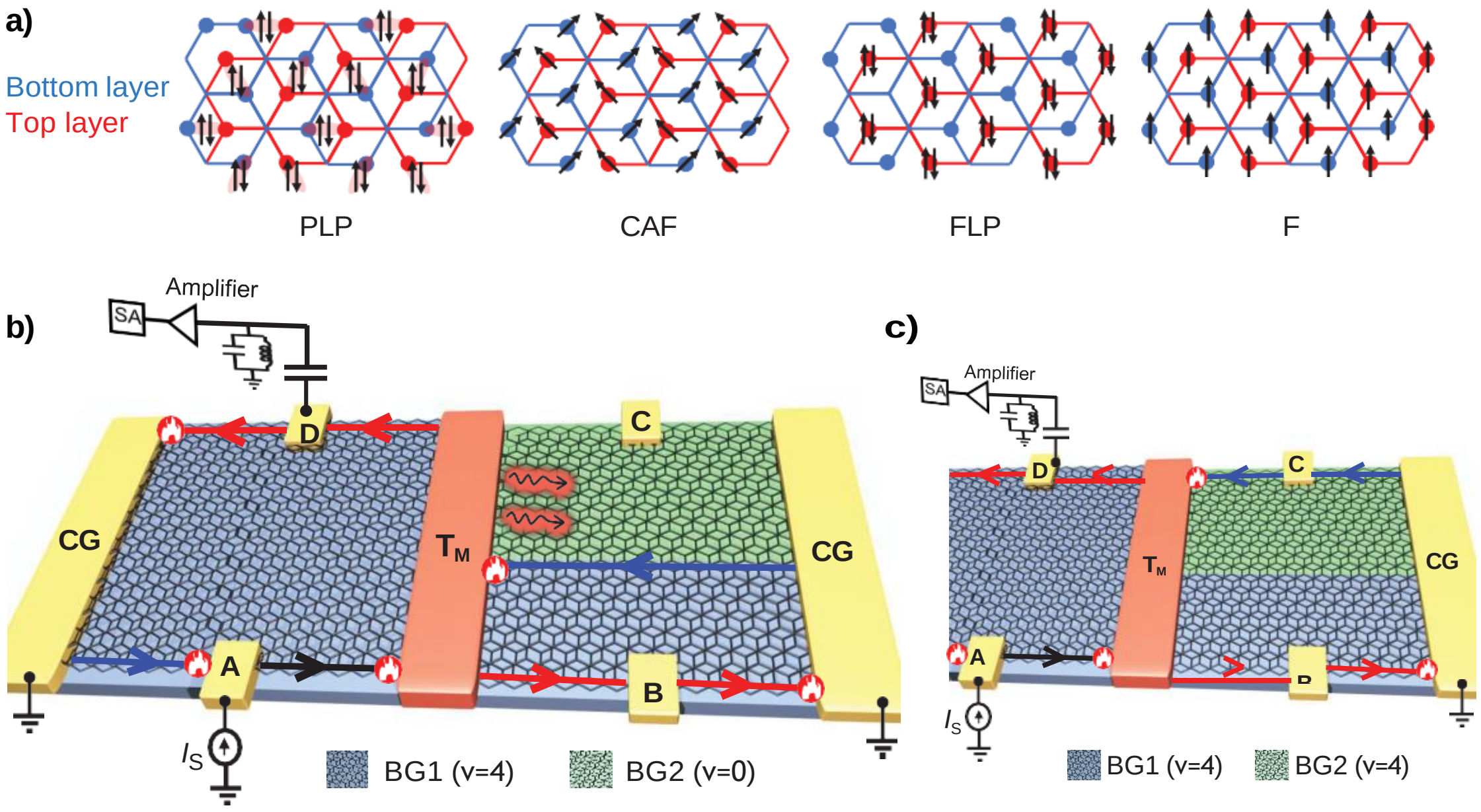

**Figure 1:** $\nu = 0$ **ground states of BLG, device and measurement scheme.** (**a**) $\nu = 0$ QH phase of BLG with four predicted ground states. (**b**) The device and measurement setup. The device is set into integer QH regime, where the global BLG part (shown in blue color and controlled by a graphite back gate - BG1) is set into $\nu = 4$ state, whereas the local BLG part (shown in green color and controlled by a local graphite back gate - BG2) is set into $\nu = 0$ state. For simplicity, four edge channels are shown by a single line with an arrow. A DC current $I_S$, is injected from contact **A**, which moves in the counter-clockwise direction set by the magnetic field. Chiral edge channels (red color) at potential $V_M = \frac{I_S}{2} \times \frac{h}{\nu e^2}$, and temperature $T_M$ leave the floating contact and terminate into two cold ground contacts. The electron temperature $T_M$ of the floating contact is determined by measuring the excess thermal noise at contact **D** at a frequency of ~730kHz using a LCR resonant circuit followed by a cascade of amplifiers and finally measured in spectrum analyzer. The wiggly line emanating from floating contact represents heat transport via Goldstone modes of $\nu = 0$ state. **c)** Same as **b)**, but here, both the global and local BLG parts are set into $\nu = 4$. The hot spots in Figures b) and c) at different corners are shown in a red circle filled with a white flame.

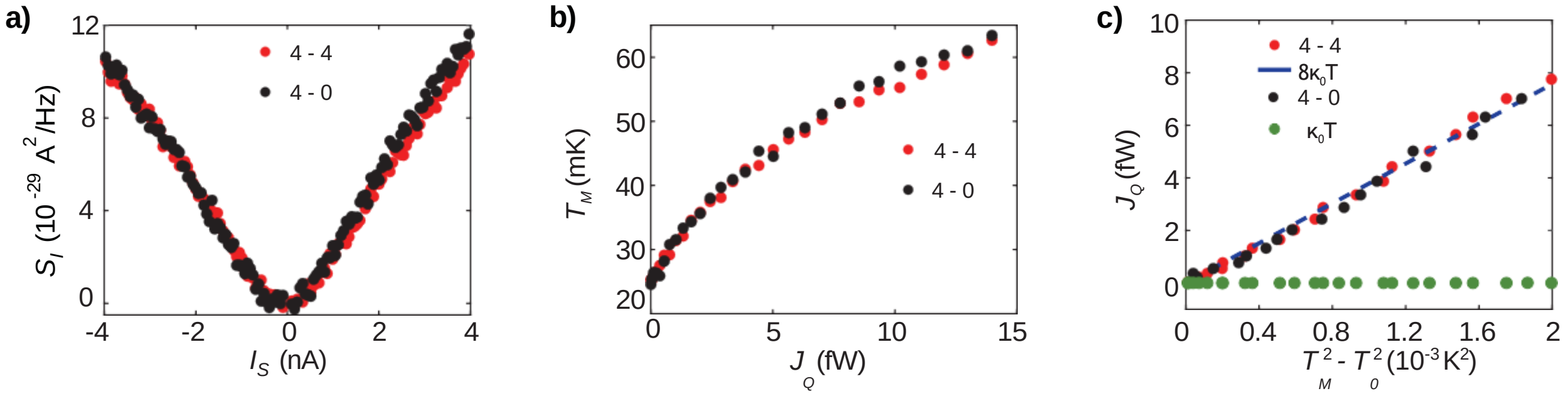

**Figure 2: Thermal conductance data at base temperature 20mK**. (**a**) Excess thermal noise $S_I$ measured as a function of the source current for device-1 for two cases: (i) $\nu_{BG1}$=4, $\nu_{BG2}$=4 (red filled circles) and (ii) $\nu_{BG1}$=4, $\nu_{BG2}$=0 (black filled circles). (**b**) The increased temperature $T_M$ of the floating contact is plotted as a function of the dissipated power $J_Q$. (**c**) $J_Q$ (solid circles) as a function of $T_M^2 - T_0^2$, and shows an excellent agreement with the $G_Q = 8\kappa_0 T$ shown by the solid dashed blue line. Green solid circles represent the difference in thermal conductance of the two cases (4,4 and 4,0) in units of $\kappa_0 T$ and show no detectable difference.

Note that we have chosen two graphite back gates in order to have the $\nu = 0$ state with zero displacement field (details in SI-S3), such that it can host the CAF ground state as theoretically suggested[30] for a moderate perpendicular magnetic field range ($B < 10T$). The local graphite back gate - BG2, is separated from the global graphite back gate - BG1, by a thin hBN spacer ($\sim 10nm$). The contacts on BLG are patterned into a Hall bar geometry with an additional metallic floating contact in the middle, which is connected to BLG via one-dimensional edge contacts (See SI-S1 for details). We have used two graphite back-gated devices: device 1 and device 2. The device 1 was cooled down to $20mK$ in a dilution refrigerator, and a perpendicular field of 4T was applied. The QH response with the robust plateaus at different filling factors can be seen in SI-Fig. 2. Electrical conductance measurements are performed with standard Lock-In techniques, whereas the thermal conductance measurement was performed with noise thermometry[35–38] based on an LC-resonant circuit, amplifiers, and a spectrum analyzer (details in SI-S4). For measuring the thermal conductance of the $\nu = 0$ state, we have performed the measurements for two different cases: (i) when both the global and the local BLG parts are at the same filling factor $\nu = 4$ (Fig. 1c), and (ii) when the global BLG part is at $\nu = 4$ while local BLG part is at $\nu = 0$ QH state (Fig. 1b). The difference in thermal conductance values of these two cases will provide the thermal conductance of the $\nu = 0$ state. If there are gapless collective excitations like Goldstone modes exist for the $\nu = 0$ state (Fig. 1b), extra heat current should flow from the hot floating contact to the cold ground in addition to the electronic heat flow via chiral edge channels, as shown schematically in Fig. 1b.

**Heat flow measurement.** We first describe the results for case (i)-without $\nu = 0$ state, where the whole BLG is set to $\nu = 4$ (Fig. 1c). A dc current ($I_S$) is injected at contact A, which flows towards the floating reservoir, and the outgoing current from the floating reservoir splits into two equal parts (see SI-Fig. 2 for details) and flows towards the cold ground contacts. The power dissipation at the floating reservoir due to hot spots (joule heating) is, $J_Q = \frac{I_S^2}{4\nu G_0}$ [35–37,39] ($G_0$ - quanta of electrical conductance), and thus the electrons in the floating reservoir reach a new steady state temperature ($T_M$) with the following heat balance relation; $J_Q = J_Q^e(T_M, T_0) = 0.5N\kappa_0(T_M^2 - T_0^2)$[35–37,39], where the $J_Q^e(T_M, T_0)$ is the electronic contribution of the heat current via $N$ chiral edge modes (here, $N = 8$ as 4 QH edge channels are leaving the floating reservoir towards the left, and 4 are leaving towards the right in Fig. 1c), and $\kappa_0 = \pi^2 k_B^2 / 3h$[35–37,39] with $k_B$ - Boltzmann constant, $h$ - Planck constant, and $T_0$ is the temperature of the cold ground. The $T_M$ of the floating reservoir is obtained by measuring the excess thermal noise at contact D given by $S_I = \nu k_B(T_M - T_0)G_0$[35–37,39], along the outgoing edge channels as shown in Fig. 1. Fig. 2a shows the measured excess thermal noise $S_I$ (red circles) as a function of current $I_S$ for case (i)-without $\nu = 0$ state (at bath temperature, $T_{bath} \sim 20mK$). The current and the noise axis of Fig. 2a are converted to $J_Q$ and $T_M$, and plotted in Fig. 2b (red circles). To extract out the value of thermal conductance, $G_Q$, we have plotted the $J_Q$ as a function of $T_M^2 - T_0^2$ in Fig. 2c. The solid red circles represent the experimental data, while the blue dashed line represents the theoretical contribution of electronic heat flow with $G_Q = 8\kappa_0 T$, where $T = (T_M + T_0)/2$. The excellent match between the experimentally measured data and theory in Fig. 2c unambiguously proves the accuracy of our measurement set-up.

Now, we will discuss the results obtained for case (ii)-with $\nu = 0$ state, where the local BLG part (BG2) is set at $\nu = 0$ while keeping the global BLG part at $\nu = 4$ (Fig. 1b). The electronic contribution to thermal conductance should remain the same as the number of electronic channels leaving the floating reservoir has not changed. However, If there are gapless collective excitations like Goldstone modes exist in the $\nu = 0$ part, the thermal conductance is expected to be higher than case (i) as $J_Q = 0.5N\kappa_0(T_M^2 - T_0^2) + J_Q^{CAF}(T_M, T_0)$, where $J_Q^{CAF}(T_M, T_0)$ is the extra heat current carried by the gapless collective excitations (as expected for the CAF phase). The wiggly lines in Fig. 1b represent the extra heat flow via gapless Goldstone modes of electrically insulating bulk of $\nu = 0$ state under the local BLG part. The results obtained for case (ii)-with $\nu = 0$ state are shown by the solid black circles in Fig. 2a, b, and c. Surprisingly, the results for case (ii)-with $\nu = 0$ state are identical to case (i)-without $\nu = 0$ state, and the measured thermal conductance matches extremely well with $8\kappa_0 T$ without any additional contribution (Fig. 2c). The measurement was repeated for device 2, allowing us to measure an additional configuration: (i) when both the global and local BLG parts are at the same filling factor, $\nu = 1$, and (ii) when the global BLG part is at $\nu = 1$ while the local BLG part is at the $\nu = 0$ state. The details are shown in SI-Fig. 4 and SI-Fig. 5, where no additional heat flow contribution is observed via $\nu = 0$ state.

With the two graphite back gates (global and local) in the above device geometry, we could not vary the $D$ to access the other phases of the $\nu = 0$ QH state of the local BLG part. To achieve this, we used the second type of device structure, device 3, which features a metallic local top gate at the top of hBN-encapsulated BLG. This structure is isolated from metallic ohmic contacts by a few-nanometer insulating layer of $Al_2O_3$, as shown schematically in Fig. 3a. Details can be found in SI-Fig. 6, where the displacement field ($D$) was varied for the local BLG part. The effect of $D$ on the $\nu = 0$ state is shown in SI-Fig. 7g, where the closing of the insulating gap of the $\nu = 0$ state is observed around $D^* \sim \pm 0.07 V/nm$. This indicates the transition from one phase of the $\nu = 0$ state to another, as reported in the literature[14, 19, 40, 41]. In Fig. 3b and 3c, we summarize the thermal conductance results for device 3 in two scenarios: (i) when both the global and local BLG parts are at the same filling factor, $\nu = 4$, and (ii) when the global BLG part is at $\nu = 4$ while the local BLG part is tuned to the $\nu = 0$ state with $D \sim -0.08$V/nm ($> D^*$). In the latter scenario, no additional contribution to heat flow for the $\nu = 0$ state was observed.

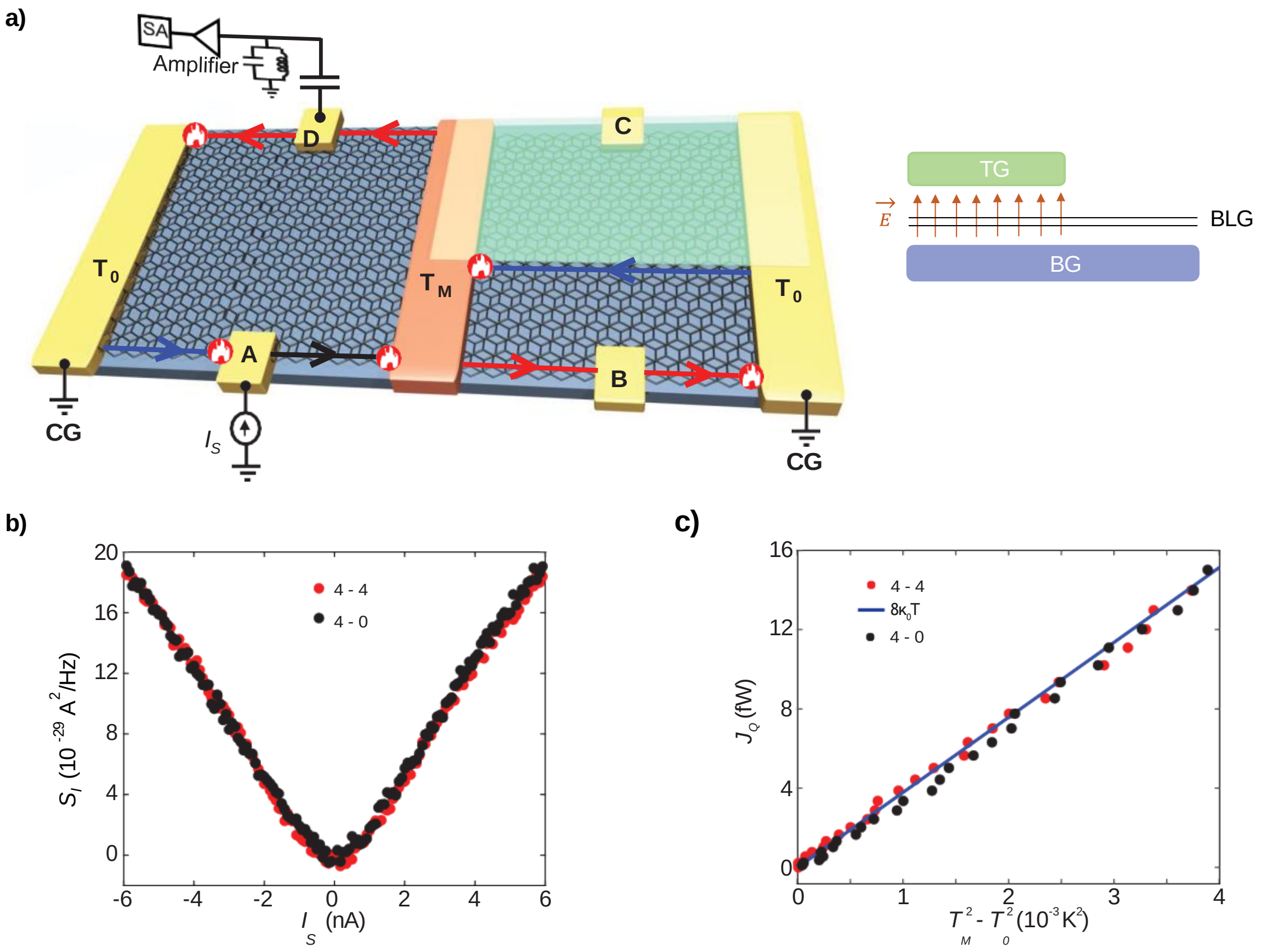

**Figure 3: Thermal transport at finite $D$. a)** Device schematic, similar to Fig. 1b, with the only difference having BG2 replaced by the top gate (TG) to apply transverse displacement field as shown schematically in the right panel. **b)** Excess thermal noise $S_I$ measured as a function of the source current for two cases: (i) $\nu_{BG}$=4, $\nu_{TG}$=4 (red) and (ii) $\nu_{BG}$=4, $\nu_{TG}$=0 (black) at $D \sim -0.08$V/nm (see SI-Fig. 7 for details). **c)** $J_Q$ (solid circles) as a function of $T_M^2 - T_0^2$. The blue solid line represents the linear fit with $G_Q = 8\kappa_0 T$.

The absence of thermal conductance at a finite $D \sim -0.08$V/nm in Fig. 3 for device 3 is expected, as the ground state of the $\nu = 0$ state is FLP with gapped excitations[1, 28, 30, 40]. However, the absence of heat flow in the $\nu = 0$ state at $D \sim 0$ for device 1 and device 2 (Fig. 2 and SI-Fig. 5) is surprising, as the $\nu = 0$ state was expected to be in the CAF phase with gapless excitations[30]. These results may suggest the following possible scenarios: **a)** either the ground states of the $\nu = 0$ state do not support the gapless collective excitations, or **b)** at the lower bath temperature ($\sim 20 mK$), the thermal conductance contribution of the collective excitations is negligible. In the next section, we discuss the possible ground states of the $\nu = 0$ state and their excitation spectra; here, we focus on the second scenario. This scenario has merit, as the contribution from collective excitations in two dimensions is expected to decrease quadratically with temperature[30]. To capture a significant contribution from the collective excitations, we conducted measurements at elevated temperatures up to $1K$ for both scenarios: (i) without $\nu = 0$ state and (ii) with $\nu = 0$ state for device 1. However, these higher temperature measurements are challenging due to potential heat loss via electron-phonon coupling from the hot metallic floating contact to the substrate, as well as heat loss via phonons of BLG from the hot metallic floating contact to the cold ground.

To simplify the analysis, we restricted our measurements to the linear regime at elevated temperatures, ensuring that $\Delta T < T_0$, where $\Delta T = (T_M - T_0)$ and $T_0$ is the temperature of the cold ground. In this regime, the heat balance equation can be simplified as follows: $J_Q = 0.5 N\kappa_0 (T_M^2 - T_0^2) + G_Q^{CAF}\Delta T + G_Q^{ph}\Delta T + G_Q^{e-ph}\Delta T$. This can be further simplified to: $J_Q = N\kappa_0 T_0 \Delta T + G_Q^{CAF}\Delta T + G_Q^{ph}\Delta T + G_Q^{e-ph}\Delta T$, where the successive terms represent contributions from electronic channels, collective excitations (CAF), phonons, and electron-phonon interactions, respectively. In Fig. 4a-c, we show the measured excess thermal noise, $S_I$, as a function of injected current, $I_S$, at $T_{bath} = 80$ mK, 300 mK, and 500 mK for both scenarios: (i) without $\nu = 0$ state and (ii) with $\nu = 0$ state. We do not observe any difference between the two cases. In Fig. 4d-f, $J_Q$ versus $\Delta T$ is plotted at $T_{bath} = 80$ mK, 300 mK, and 500 mK for one of the scenarios. It can be seen that for a small $\Delta T$ range ($\Delta T < T_0$), $J_Q$ increases linearly with $\Delta T$, and the slope gives the value of the thermal conductance, indicated by the solid blue lines. The solid black lines show the expected contribution from the electronic edge channels ($N\kappa_0 T \simeq 8\kappa_0 T_0$). Unlike at low temperatures (Fig. 2), at elevated temperatures, the measured thermal conductance is much higher than the electronic part, as expected from additional contributions. However, the absence of any detectable difference between the two scenarios (i) without $\nu = 0$ state and (ii) with $\nu = 0$ state in Fig. 4a-c suggests that there is no heat flow via collective excitations of isospins in the $\nu = 0$ quantum Hall ferromagnet of bilayer graphene. Note that the cooling of the electron system remains efficient down to our lowest bath temperature. The electron temperature, $T_0$ remains very close to $T_{bath}$ throughout the entire temperature range of our measurements. At the lowest $T_{bath} \sim 20mK$, the $T_0$ is $\sim 24mK$. Details are shown in SI-S5 and can be found in our previous work[35, 37, 38].

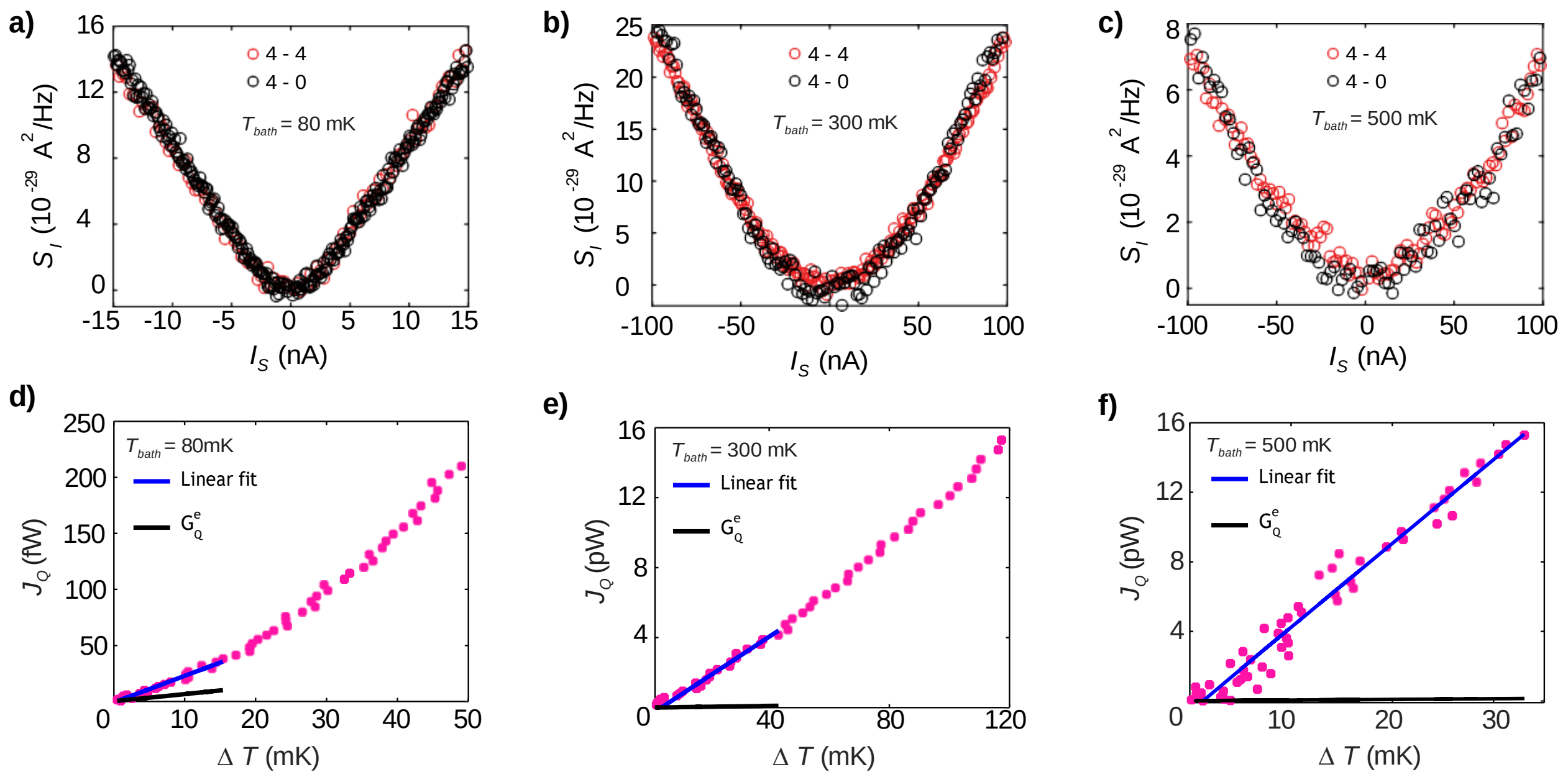

**Figure 4: Thermal conductance data at higher temperatures**: Excess thermal noise $S_I$ measured as a function of the source current at temperatures 80mK (**a**), 300mK (**b**), and 500mK (**c**) for device-1 for two cases (i) $\nu_{BG1}$=4, $\nu_{BG2}$=4 (red), and (ii) $\nu_{BG1}$=4, $\nu_{BG2}$=0 (black). For the case (ii) (4-0), $J_Q$ vs $\Delta T$ plotted for 80mK (**d**), 300mK (**e**) and 500mK (**f**). To calculate the thermal conductance, the data is fitted linearly (solid blue line) in a small $\Delta T$ range. The solid black line represents the expected electronic contribution of thermal conductance.

To elucidate our findings at elevated temperatures, as depicted in Fig. 4, we aimed to analyze the two-dimensional heat current flow through collective excitations. At lower temperatures, it is reasonable to assume that heat transport occurs ballistically through collective excitations of isospin or phonons[30]. For linearly dispersive excitations, the thermal conductance value can be expressed as approximately $\sim WT^2/v$ (Ref[30]), where $W$ and $v$ represent the width of the device and the velocity of the excitations, respectively. In Fig. 5, we have plotted the additional contribution to thermal conductance (extracted from the measured total thermal conductance in Fig. 4, excluding the electronic part) as a function of the bath temperature using solid circles. The data is well-fitted with $2.7 \times 10^{-9} \times T^{2.5}$, indicated by the solid black line. The power exponent of around $\sim 2$ implies that the additional contribution to thermal conductance arises from collective excitations. However, the absence of a discernible difference between scenarios i)-without $\nu = 0$ state and ii)-with $\nu = 0$ state suggests that the primary origin of the additional contribution at the elevated temperature is from phonons ($G_Q^{ph}$). Note that the thermal conductance from ballistic phonons is expected to exhibit a $T^2$ dependence. However, the observed power exponent greater than *2* suggests the influence of electron-phonon coupling ($G_Q^{e\text{–}ph}$), which facilitates heat loss from the hot electrons of the floating metal to the phonons within the metal. The $G_Q^{e\text{–}ph}$ is anticipated to have a higher power exponent ($T^{\alpha}$, where $\alpha \sim 4 - 5$) [35]. Additionally, the quadratically dispersive flexural modes of graphene can result in different power dependencies. Future studies are necessary to fully understand the power exponent at elevated temperatures.

**Discussion.** According to Ref. [1], the four ground states of the $\nu = 0$ state are CAF, PLP, FLP, and F. Among these, the excitations remain gapped for FLP and F, while CAF and PLP are expected to support gapless collective excitations, as suggested by Ref. [30]. Notably, around $T_0 \sim 100$ mK, the thermal conductance from the gapless collective excitations is nearly twice that of an electronic edge channel [30] (see SI-S8). Therefore, the absence of any thermal conductance of the $\nu = 0$ state in device 1 and device 2 suggests the following scenarios (details in SI-S6 and SI-S7):

(i) The ground state could be either FLP or F; however, the electrically insulating nature of the $\nu = 0$ state rules out the F phase. The FLP phase is also unlikely, as the measurements were conducted at zero displacement field, and the BLG was not aligned with the hBN substrate [1, 30], as indicated by the temperature-dependent resistance data (see SI-Fig. 3).

(ii) The ground state could be CAF, which is the most expected phase for our devices based on existing transport experiments [14, 28, 40] and theories [1, 30]. However, recent STM experiments [29, 42, 43] suggest that the phase could be PLP or even a coexistence of CAF+PLP, as suggested by recent theories [5]. Although Ref. [30] indicates that collective excitations for the PLP phase are gapless, the consideration of higher-order in-teractions is expected to break the C3 symmetry. Consequently, the PLP phase may no longer remain a continuous symmetry-breaking phase capable of exhibiting gapless Goldstone modes [5, 13, 44]. For the same reason, a gapped spectrum is also expected for the coexistence CAF+PLP phase [5].

(iii) If the ground state is the CAF phase, gapless excitations are anticipated due to spontaneously broken continuous symmetry. However, the spectrum of CAF may be gapped due to the quantization of collective modes arising from the finite size of the device, an effect observed in BLG devices [28]. The upper limit of the gap from this confinement effect could be as significant as 350-650 mK for our device dimensions (see SI-S8 for details).

(iv) There might be insufficient coupling between the hot electron bath of the metal and the collective ex-citations of BLG. However, this is unlikely for our devices, as the metal contacts tend to locally n-dope the BLG flake [45–47], thus the hot electrons of the reservoir extend into the BLG, potentially enhancing the cou-pling to collective excitations. For instance, the local n-doping confines the edge channels near the contacts, and the tunnelling of electrons between edge channels with opposite spin can generate and absorb collective excitations. The details can be found in SI-S9.

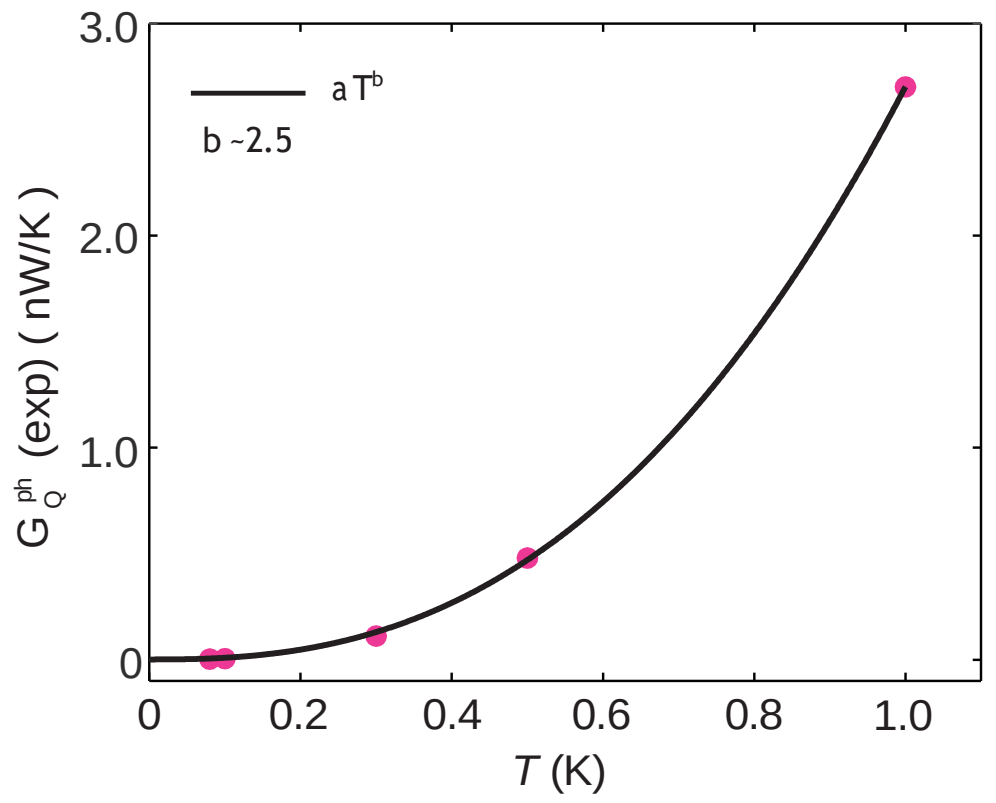

**Figure 5: Contribution of phonon.** Additional contribution of thermal conductance as a function of bath temperature shown in red solid circles. The solid black line is the best fitting with $\sim T^{2.5}$.

Therefore, it is most likely that the gap in the excitation spectra of the PLP phase (due to C3 symmetry breaking) or the CAF phase (due to quantization from finite size) inhibits heat transport, resulting in the absence of thermal conductance at lower temperatures. At elevated temperatures, these gaps could be overcome. However, at higher temperatures, phonons become dominant, as illustrated in Fig. 5. At 1K, the phonon contribution is more than an order of magnitude greater than the expected contribution from the collective excitations of the CAF phase [30]. A similar experiment conducted in monolayer graphene [48] also showed vanishing bulk heat flow at the $\nu = 0$ state, suggesting an analogy between monolayer and bilayer graphene. These findings point to potential commonalities in the absence of heat flow at the $\nu = 0$ state in both MLG and BLG, highlighting the need for further theoretical and experimental efforts to fully understand the nature of excitations in the $\nu = 0$ states of MLG and BLG.

**Conclusion.** In conclusion, our study explored thermal transport measurements of the $\nu = 0$ quantum Hall ferromagnet state in bilayer graphene by varying the bath temperature under different displacement fields. The absence of a notable contribution to thermal conductance from the $\nu = 0$ state at zero displacement field suggests the presence of gapped excitations in the device. This work demonstrates the importance of thermal transport measurement to explore electrically insulating ground states and their collective excitations in bulk phases within the integer and the fractional quantum Hall regime.

## Materials and Methods

**Device fabrication and measurement scheme:** In our experiment, an encapsulated device (heterostructure of hBN/bilayer graphene(BLG)/hBN/graphite/hBN/graphite) was made using the standard dry transfer pick-up technique [49]. Fabrication of this heterostructure involved mechanical exfoliation of hBN and graphite crystals on an oxidized silicon wafer using the widely used scotch tape technique. First, a hBN of the thickness of $\sim 25-35$ nm was picked up at $90^\circ$C using a Poly-Bisphenol-A-Carbonate (PC) coated Polydimethylsiloxane (PDMS) stamp placed on a glass slide, attached to the tip of a home built micro-manipulator. This hBN flake was placed on top of the previously exfoliated BLG. BLG was picked up at $90^\circ$C. The next step involved picking up the bottom hBN ($\sim 30-50$ nm). Following the previous process, this bottom hBN was picked up using the previously picked-up hBN/BLG. Following the previous step, this hBN/BLG/hBN heterostructure was used to pick up the graphite flake (working as the local gate in the device). Now, this heterostructure (hBN/BLG/hBN/graphite) was used to pick up another hBN flake and followed by pick up of another graphite flake (works as global gate). Finally, the heterostructure was dropped down on top of an oxidized silicon wafer of thickness 285 nm at temperature $180^\circ$C. To remove the residues of PC, this final stack was cleaned in chloroform ($CHCl_3$) overnight followed by cleaning in acetone and isopropyl alcohol (IPA). After this, Poly-methyl-methacrylate (PMMA) photoresist was coated on this heterostructure to define the contact regions in the Hall probe geometry using electron beam lithography (EBL). Apart from the conventional Hall probe geometry, we defined a region of $\sim$ 5.5 $\mu m^2$ area in the middle of BLG flake, which acts as a floating metallic reservoir upon edge contact metallization. After EBL, reactive ion etching (mixture of $CHF_3$ and $O_2$ gas with a flow rate of 40 sccm and 4 sccm, respectively at $25^\circ$C with RF power of 60W) was used to define the edge contact[50]. The etching time was optimized such that the bottom hBN did not etch completely to isolate the contacts from graphite flakes, which were used as gates. Finally, thermal deposition of Cr/Pd/Au (4/12/70 nm) was done in an evaporator chamber having base pressure of $\sim 1-2\times 10^{-7}$ mbar. After deposition, a lift-off procedure was performed in hot acetone and IPA. This results in a Hall bar device and the floating metallic reservoir connected to both sides of BLG by the edge contacts. The device's schematics and measurement setup are shown in Fig. 1(b). The distance from the floating contact to the ground contacts was $\sim 5\mu m$ (see SI-S1 for optical images).

For the device structure with a metallic top gate, a hBN/BLG/hBN/graphite heterostructure was created. Standard Hall bar geometry with a metallic floating contact in the middle was fabricated. At last, 5nm thick oxide layer of $Al_2O_3$ was evaporated on the device followed by deposition of aluminum metal, working as the local top gate. All measurements were done in a cryo-free dilution refrigerator having a base temperature of $\sim 20$mK. The electrical conductance was measured using the standard lock-in technique, whereas the thermal conductance was measured with noise thermometry based on an LCR resonant circuit at resonance frequency $\sim 725$kHz. The signal was amplified by a home-made preamplifier at 4K followed by a room temperature amplifier, and finally measured by a spectrum analyzer. Details of measurement technique are discussed in our previous work [35–37] and in the SI-S4 and SI-S5. The QH responses of the three devices are shown in SI-S2.

**Acknowledgements**

The authors thank Dr. Ankur Das for useful discussions. A.D. thanks the Department of Science and Technology (DST) and Science and Engineering Research Board (SERB), India, for financial support (SP/SERB-22-0387) and acknowledges the Swarnajayanti Fellowship of the DST/SJF/PSA-03/2018-19. A.D. also thanks financial support from CEFIPRA: SP/IFCP-22-0005. K.W. and T.T. acknowledge support from the Elemental Strategy Initiative conducted by the MEXT, Japan and the CREST (JPMJCR15F3), JST.

**Author contributions**

R.K., S.K.S., and U.R. contributed to device fabrication, data acquisition, and analysis. A.D. contributed to conceiving the idea and designing the experiment, data interpretation, and analysis. K.W. and T.T. synthesized the hBN single crystals. U.S. and V.S. contributed to $Al_2O_3$ film deposition and discussions. P.R. contributed to the discussions and data interpretation. All the authors contributed to writing the manuscript.

**Competing interests**

The authors declare no competing interests.

**Data and materials availability:**

The data presented in the manuscript are available from the corresponding author upon request.

# Supplementary information: Absence of heat flow in $\nu = 0$ quantum Hall ferromagnet in bilayer graphene

Ravi Kumar[1,*], Saurabh Kumar Srivastav[1,*], Ujjal Roy[1,*], Ujjawal Singhal[1], K. Watanabe[2], T. Taniguchi[2], Vibhor Singh[1], P. Roulleau[3] and Anindya Das[1†]

[1]*Department of Physics, Indian Institute of Science, Bangalore, 560012, India.*

[2]*National Institute of Material Science, 1-1 Namiki, Tsukuba 305-0044, Japan.*

[3]*SPEC, CEA, CNRS, Université Paris-Saclay, CEA Saclay, 91191 Gif sur Yvette Cedex, France.*

[*]These authors contributed equally: Ravi Kumar, Saurabh Kumar Srivastav, Ujjal Roy

[†]anindya@iisc.ac.in

**This supplementary information contains the following details:**

**1. Different kinds of devices for measuring thermal conductance**

**2. Device characterization**

**3. Estimation of displacement fields**

**4. Noise measurement setup**

**5. Gain and electron temperature estimation**

7. **Phase diagram of $\nu = 0$ state of BLG**

8. **Expected thermal conductance response from different ground states**

**6. Estimations of the gap of collective excitations**

**8. Thermal interface resistance**

## Section S1: Different kinds of devices for measuring thermal conductance

We have used three devices in our measurements: device 1 and device 2 are used to measure the thermal conductance at zero displacement field, and device 3 is used to measure at finite displacement field. Supplementary Fig. 1 shows the optical image of the Device 1. Device 1 consists of a hBN encapsulated BLG placed on two independent graphite back gates, a global graphite back gate - BG, and a local graphite back gate - LG, to control the charge carrier density or the filling factors. The local graphite back gate - LG, is separated from the global graphite back gate - BG, by a thin hBN spacer ($\sim 10nm$). The contacts on BLG are patterned into a Hall bar geometry with an additional metallic floating contact in the middle, which is connected to BLG via one-dimensional edge contacts, as shown in supplementary Fig. 1. For measuring the thermal conductance of the CAF phase with Goldstone modes, we have performed the measurements for two different cases: (i) when both the global and the local BLG parts are at the same filling factor $\nu = 4$, and (ii) when the global BLG part is at $\nu = 4$ but local BLG part is at $\nu = 0$ quantum Hall (QH) state. The BG and LG independently control the states $\nu = 0$ and $\nu = 4$ of the local and global parts. Note that the LG screens the BG in this geometry. Similar to device 1, device 2 is also controlled by the bottom LG and BG; however, in this device, we could measure for the additional configuration with (i) when both the global and the local BLG parts are at the same filling factor $\nu = 1$, and (ii) when the global BLG part is at $\nu = 1$, but local BLG part is at $\nu = 0$ quantum Hall (QH) state.

However, With the two graphite back gates (global and local) in device 1 and device 2, one can not vary the displacement field, $D$, to access the other phases of the $\nu = 0$ QH state of BLG. To do so, we have used the third device, device 3, with a metallic local top gate at the top of hBN-encapsulated BLG isolated from metallic ohmic contacts via an insulating $Al_2O_3$ layer of few nm as shown in supplementary Fig. 1b. In device 3, the BG controls the charge carrier density or the filling factors of the global part, but in the local part, the carrier density depends on both the BG and TG. In this device, one can independently control the carrier density and displacement field of the local region.

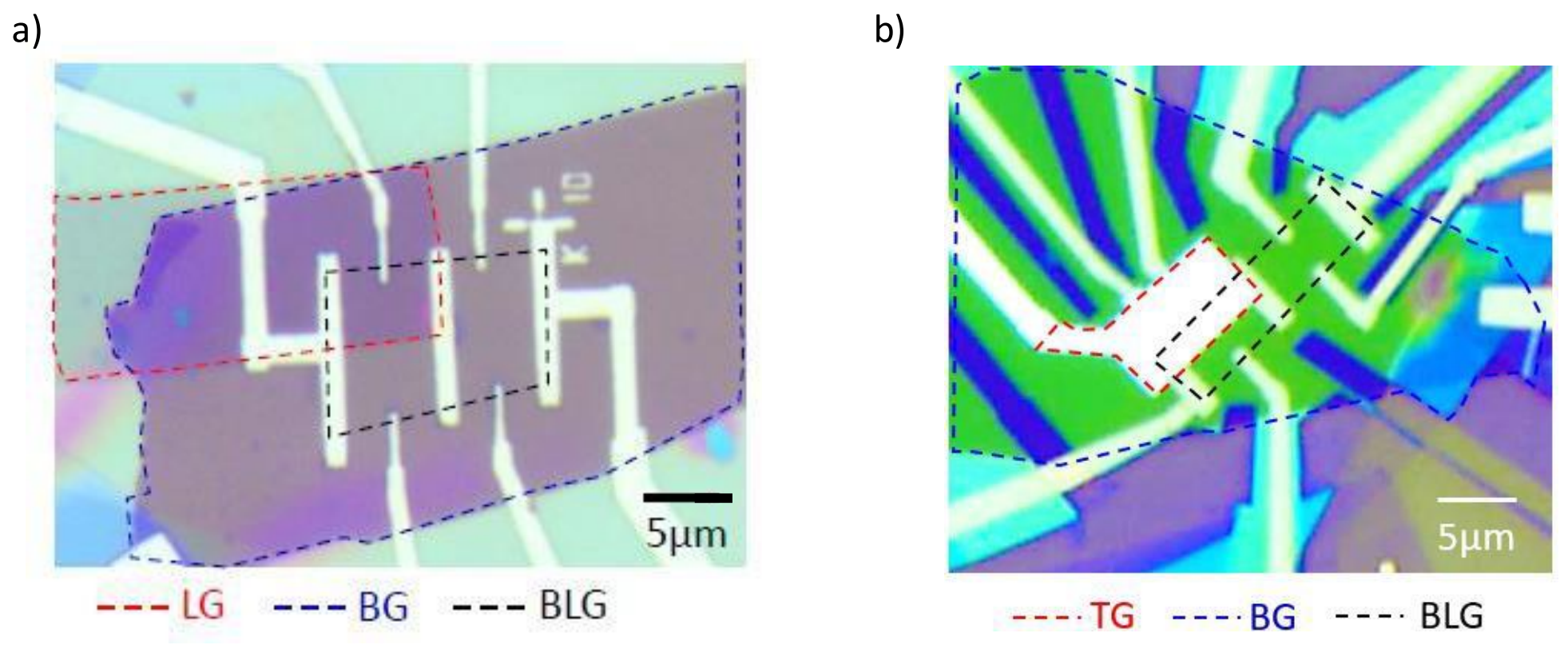

**Supplementary Figure 1: Optical image of the devices** (**a**) Optical image of the device 1. BLG channel, local back-gate, and global back-gate are outlined by dashed black, red, and blue lines. The geometry remains the same for device 2. (**b**) Optical image of the device 3. Dashed black, red, and blue lines outline BLG channel, aluminum local top-gate, and global back-gate.

**Section S2: Device characterization**

The devices were characterized at zero and finite magnetic fields in the quantum Hall regime. This was achieved by measuring resistance as a function of the gate voltages (LG and BG) with a standard Lock-In technique at a low frequency of 23 Hz. The characterization of device 1 (with local back gate) is shown in supplementary Fig. 2 for B=0T and B=4T. The robust QH plateaus are seen for device 1 at 4T. Furthermore, the equipartition of the current from the floating contact is established when the fillings at LG and BG are the same ($\nu = 4$) and when the filling at LG is zero ($\nu_{BG} = 4$, $\nu_{LG} = 0$, supplementary Fig. 2f). However, for this device the equipartition could not be seen at $\nu = 1$. QH plateaus of device 1 remain robust at higher temperatures, as shown in supplementary Fig. 3-right panel. This allows us to do thermal conductance measurements at higher temperatures. It should also be noted that the resistance of the Dirac point measured at contact A under the LG part hardly changes with temperature from 20mK to 2K (supplementary Fig. 3 - left panel), suggesting lack of any gap formation in device 1, which confirms the BLG is not aligned with hBN. The supplementary Fig. 4 shows the characterization of device 2. QH plateaus at B = 8T remain robust for this device, and equipartition is seen for both the $\nu = 4$ and $\nu = 1$ fillings. For this device, the resistance value (similar to device 1) of the Dirac point at 20 mK suggests no gap opening. The thermal conductance for device 2 is summarized in supplementary Fig. 5, and no detectable contribution for $\nu = 0$ was measured. The scaling of thermal conductance with the number of edge states is shown in supplementary Fig. 5g.

The characterization of device 3 (with local top gate) is shown at B=0T and B=6T, respectively, in supplementary Fig 6 and Fig 7. The band gap opening at the Dirac point under the top gate region can be seen with the application of displacement field at zero magnetic fields (supplementary Fig. 6b). At B = 6T, the gap closing of the $\nu = 0$ could be seen around $D* \sim \pm 0.07 V/nm$ (supplementary Fig. 7g) indicating transition from the expected CAF to FLP phase as reported in literature[1].

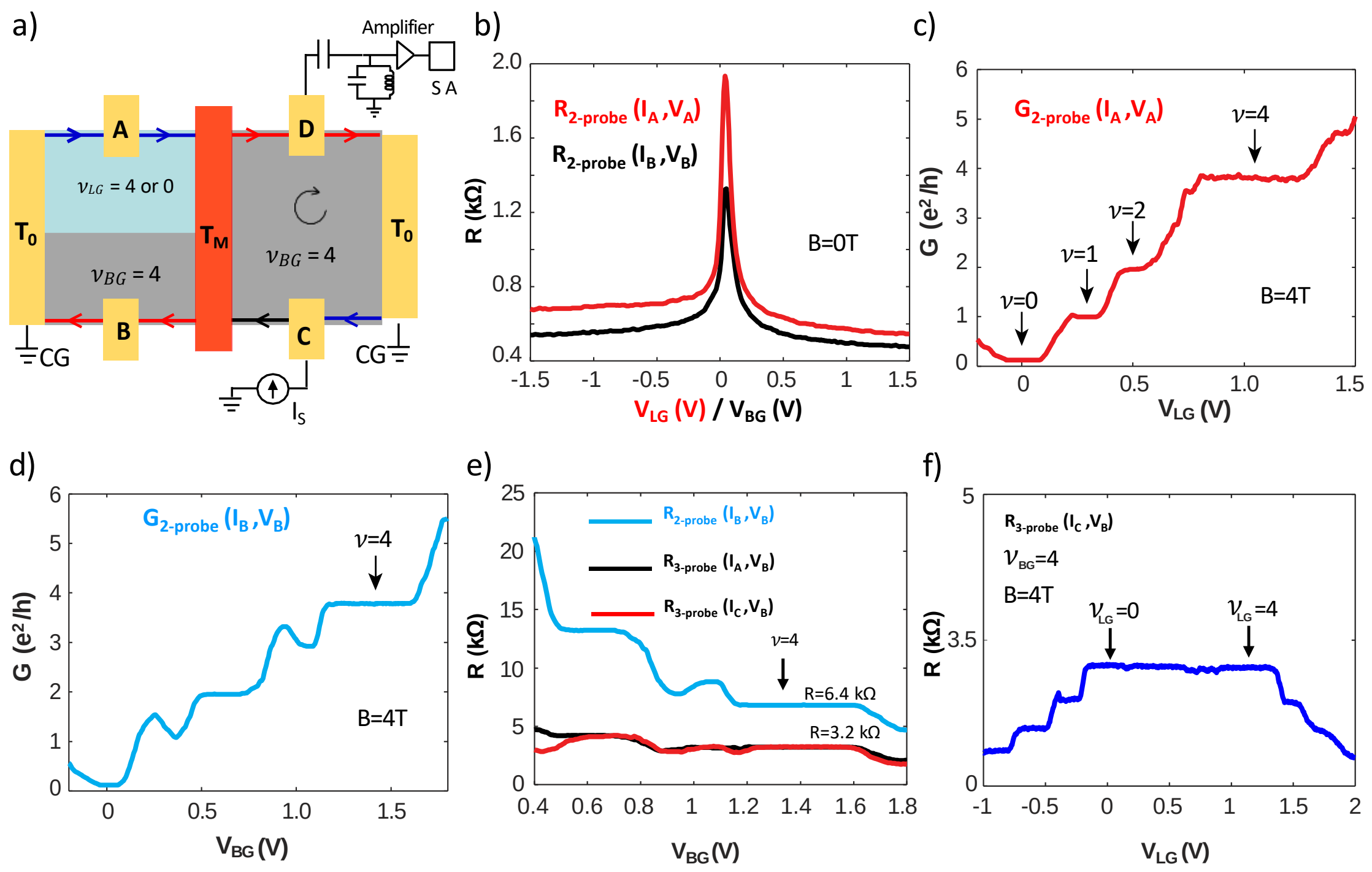

**Supplementary Figure 2: Quantum Hall response of device 1.** (**a**) Schematic of the measurement setup to characterize the device. The global back gate controls the BLG shown in grey color, and the sky blue color represents the BLG controlled by the local back gate. (**b**) Two probe resistance of the BLG as a function of local back gate (red) and global back gate (black) at B=0T. The subscripts in I and V represent current feed contact and contact where voltage is measured, respectively. (**c**) Quantum Hall response of the BLG controlled by the local back gate at $B = 4T$ . Clear plateaus developed at $\nu = 1$, 2 and 4. The region near the charge neutrality point belongs to the highly insulating $\nu = 0$ state. (**d**) Quantum Hall response of the BLG controlled by the global back gate at $B = 4T$ . The narrow peak and dip-like structures arise due to two probe measurement configurations as reported in the literature [2], which depends on the details of the device aspect ratio. Further, these structures appear around $\nu = 3$ and $\nu = 1$ plateau, which are not well developed at B = 4T. (**e**) Cyan color data shows the 2-probe QH response of contact B. The same data in Fig. d is plotted in resistance. The vertical arrow indicates the $\nu = 4$ QH plateau. Red data is the 3-probe QH response where the current is injected at contact C, and voltage is measured on the transmitted side at contact B. Similarly, black color data shows the 3-probe resistance, where the current is injected at contact A with local filling $\nu = 4$, and voltage is measured at the reflected side at contact B. The 3-probe resistance measured in both cases is exactly half of the 2-probe resistance measured for $I_B$, $V_B$ configuration for the $\nu = 4$, confirming equal current partitioning at the floating reservoir. (**f**) 3-probe resistance ($I_c$, $V_B$) as a function of $V_{LG}$ while $\nu_{BG} = 4$. The plateau appearing at the half of the quantum resistance of $\nu = 4$ at both the $\nu_{LG} = 4$ and $\nu_{LG} = 0$ (indicated by the vertical arrows) reconfirms the equipartition of the current from the floating reservoir for both the ($\nu_{BG} = 4$, $\nu_{LG} = 4$) and ($\nu_{BG} = 4$, $\nu_{LG} = 0$) cases. From Fig. 2b and 2c, it can be seen that the Dirac point or the charge neutrality point under the local back gate part appears at $V_{LG} \sim 0$V.

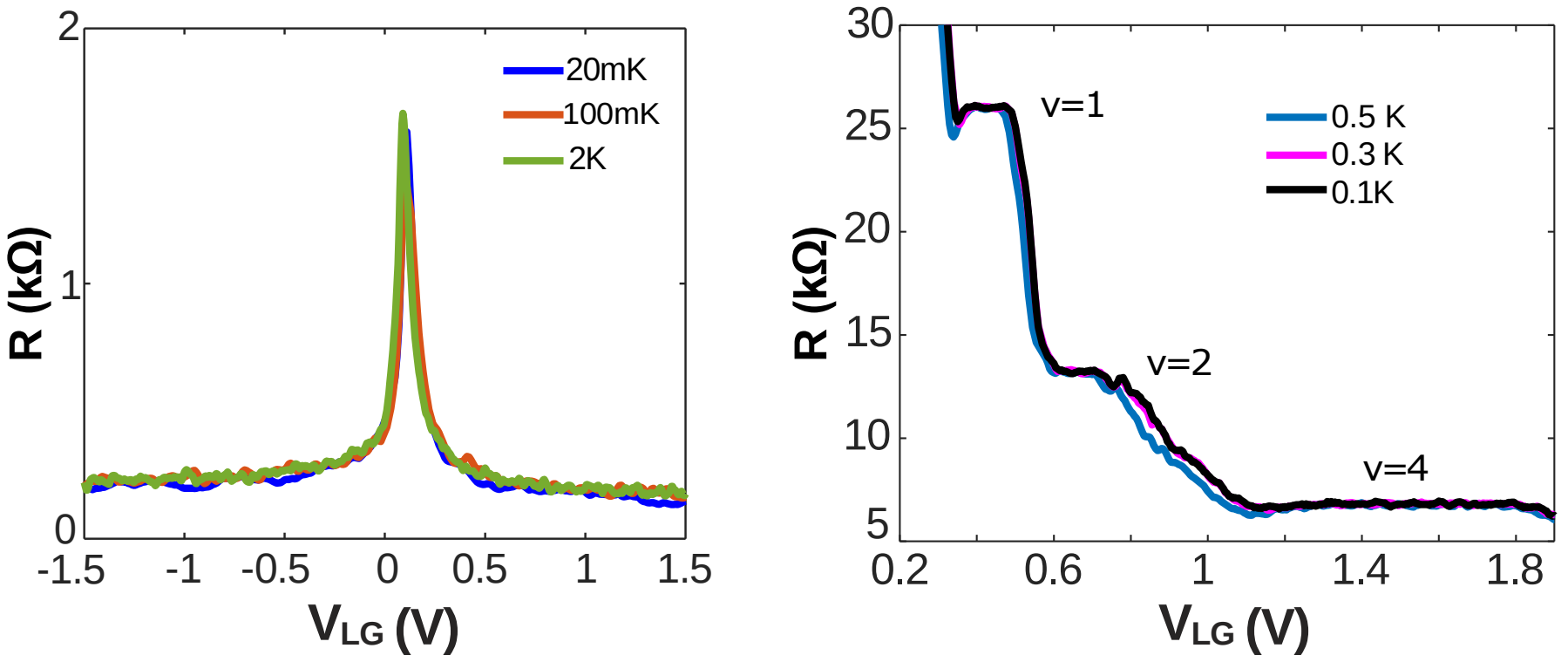

**Supplementary Figure 3:** Left panel-Two probe resistances of the BLG as a function of the local back gate voltage ($I_A$, $V_A$) at 20 mK, 100 mK, and 1K for device 1. The resistance value at the Dirac point hardly changes within this temperature range, indicating no gap opening. Right panel - QH response of device-1 at higher temperatures as a function of local gate recorded at $T \sim 0.1$ K to 0.5 K, and it can be seen the plateau remains robust at higher temperatures.

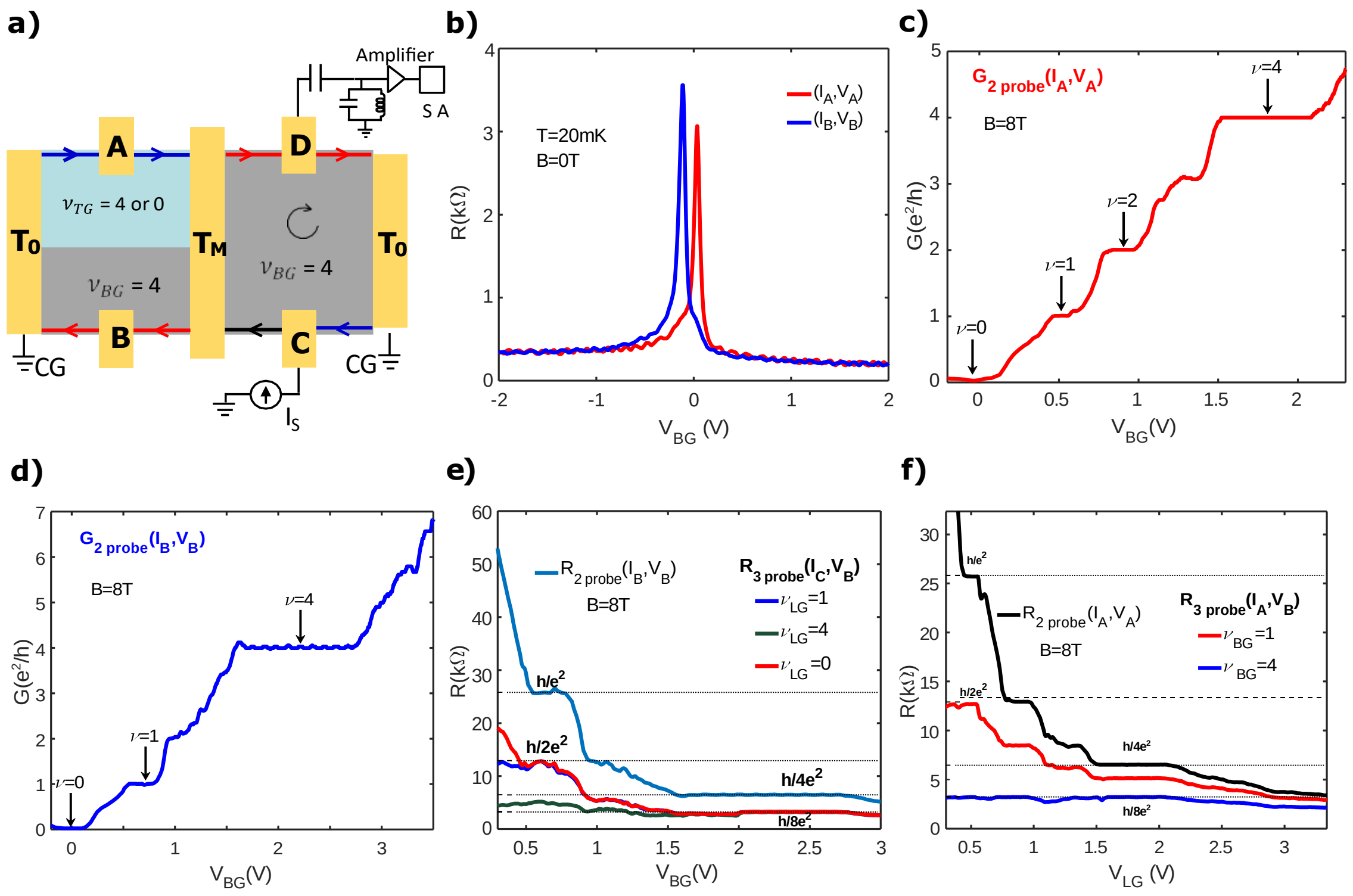

**Supplementary Figure 4: Quantum Hall response of device 2.** (**a**) Schematic of the measurement setup to characterize device 2. The global back gate controls the BLG shown in grey color, and the sky blue color represents the BLG controlled by the local back gate. (**b**) Two probe resistance of the BLG as a function of local back gate (red) and global back gate (blue) at B=0T. The subscripts in I and V represent current feed contact and contact where voltage is measured, respectively. (**c**) Quantum Hall response of the BLG controlled by the local back gate at $B = 8T$. Clear plateaus developed at $\nu = 1$, 2 and 4. The region near the charge neutrality point belongs to the highly insulating $\nu = 0$ state. (**d**) Quantum Hall response of the BLG controlled by the global back gate at $B = 8T$. (**e**) Cyan color data shows the 2-probe QH response of contact B. The other colors represent the 3-probe QH response, where the current is injected at contact C, and voltage is measured on the transmitted side at contact B as a function $V_{BG}$ for keeping the $\nu_{LG}$ at 4 (olive), 1 (blue) and 0 (red), respectively. The combination of ($\nu_{BG} = 4$, $\nu_{LG} = 4$) and ($\nu_{BG} = 4$, $\nu_{LG} = 0$) as well as ($\nu_{BG} = 1$, $\nu_{LG} = 1$) and ($\nu_{BG} = 1$, $\nu_{LG} = 0$) shows the expected half of the quantum resistance of $\nu = 4$ and $\nu = 1$, respectively, (marked by the dashed horizontal lines). This confirms the equipartition of the current from the floating reservoir for both the $\nu = 4$ and $\nu = 1$ fillings. (**f**) The equipartition is further confirmed in 3-probe QH response while injecting current at A and measuring voltage at B. From Fig. 4b and 4c, it can be seen that the Dirac point or the charge neutrality point under the local back gate part appears at $V_{LG} \sim 0$V. The resistance value (at 20 mK) at the Dirac point under the local gate (Fig. 4b - red line) is similar in magnitude to device 1 and indicates no gap opening. The beating patterns in some figures (Fig. 4b and Fig. 4d) arise due to extrinsic origins, like interference between the measuring frequency and data acquisition rate.

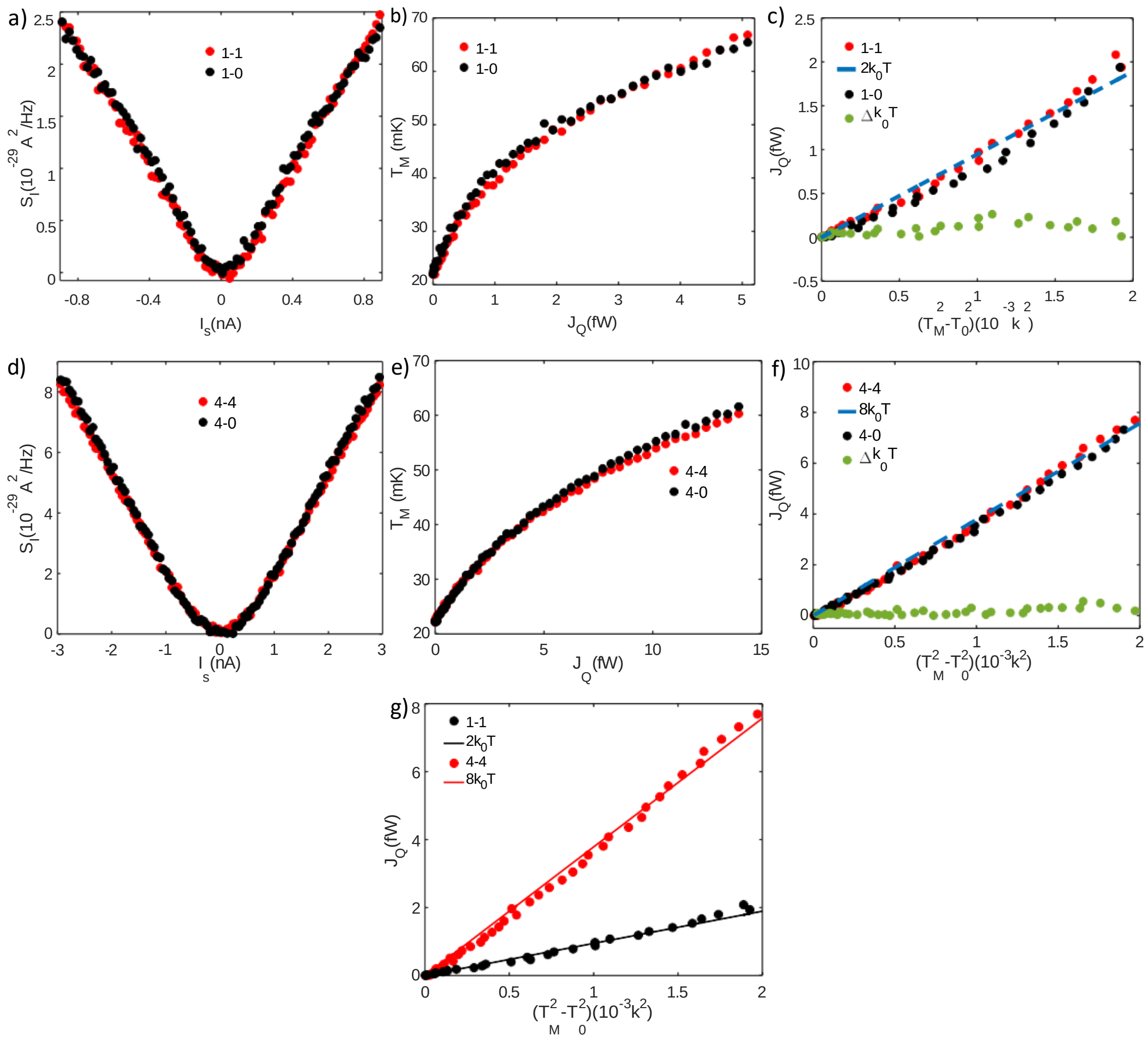

**Supplementary Figure 5: Thermal conductance data for device 2**. (**a**) Excess thermal noise $S_I$ measured as a function of the source current measured at base temperature $\sim 20mK$ for two cases: (i) $\nu_{BG}$=1, $\nu_{LG}$=1 (red) and (ii) $\nu_{BG}$=1, $\nu_{LG}$=0 (black). The increased temperature $T_M$ of the floating contact is plotted as a function of the dissipated power $J_Q$ in (**b**). (**c**) $J_Q$ (solid circles) as a function of $T_M^2 - T_0^2$. The blue dashed line represents the linear fit with $G_Q = 2\kappa_0 T$ . Green solid circles represent the difference in thermal conductance of the two cases (1,1 and 1,0) in units of $\kappa_0 T$ . No detectable thermal conductance contribution was measured for the $\nu = 0$ state. (**d**), (**e**) and (**f**) represent similar measurements with combinations of ($\nu_{BG}$=4, $\nu_{LG}$=4) and ($\nu_{BG}$=4, $\nu_{LG}$=0), and no detectable thermal conductance contribution was measured for the $\nu = 0$ state. (**g**) Comparison of thermal conductances for ($\nu_{BG}$=1, $\nu_{LG}$=1) and ($\nu_{BG}$=4, $\nu_{LG}$=4) and matches well with the expected theoretical contributions (the solid lines). This establishes that the measured thermal conductance scales with the number of edge states ($N$) in our device.

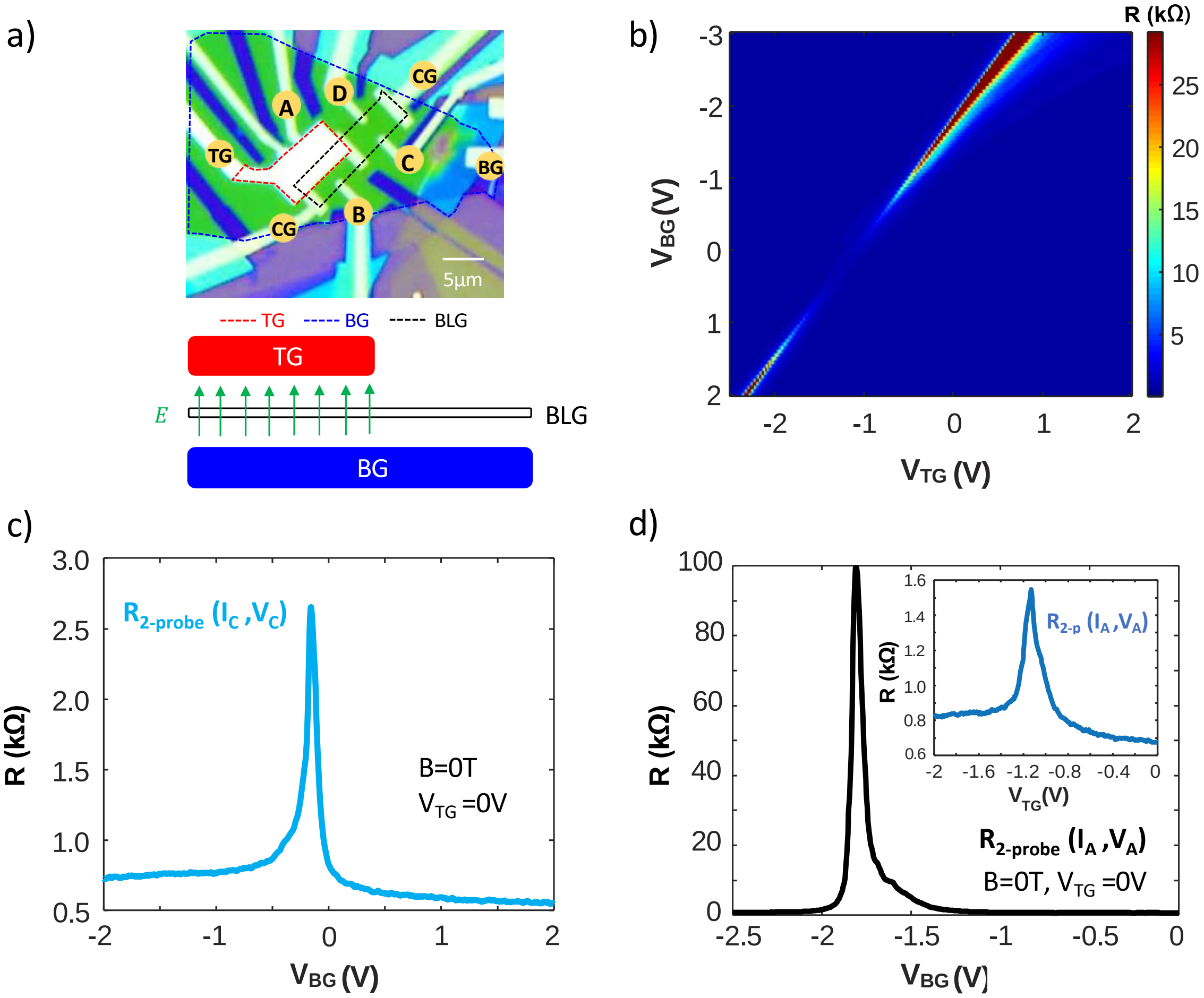

**Supplementary Figure 6: Device characterization of the device 3 at zero magnetic field.** (**a**) Optical image of the device 3. BLG channel, aluminium local top-gate, and global back-gate are outlined by dashed black, red, and blue lines, respectively. The bottom inset shows the schematic illustration of the transverse electric field applied to bilayer graphene by the combination of local top-gate and global back-gate. **(b)** 2D color map of the resistance measured (B = 0T) under the local top gate ($I_A$, $V_A$) as a function of the global back gate and local top gate voltages. The resistance value along the charge neutrality point increases with the electric field indicating the expected band gap opening in bilayer graphene. Note that the color bar is saturated to see the intrinsic positions of the Dirac points. **(c)** Two-probe resistance as a function of the global back-gate, measured at contacts C for $V_{TG}$ = 0V. The subscripts in I and V represent current feed contact and contact where voltage is measured, respectively. **(d)** Two-probe resistance as a function of the global back-gate, measured at contact A for $V_{TG}$ = 0V. The inset shows the two-probe resistance of contact A as a function of the local top gate at $V_{BG}$=0V. It should be noticed that the Dirac point under the local top gate appears at a sufficiently larger negative back gate voltage ($\sim -1.8$V) in Fig. 6d compared to the Dirac point of the global part ($\sim -0.2$V) in Fig. 6c. This indicates the presence of unintentional doping in the local part under the top gate, most likely due to static charges in the $Al_2O3$ layer.

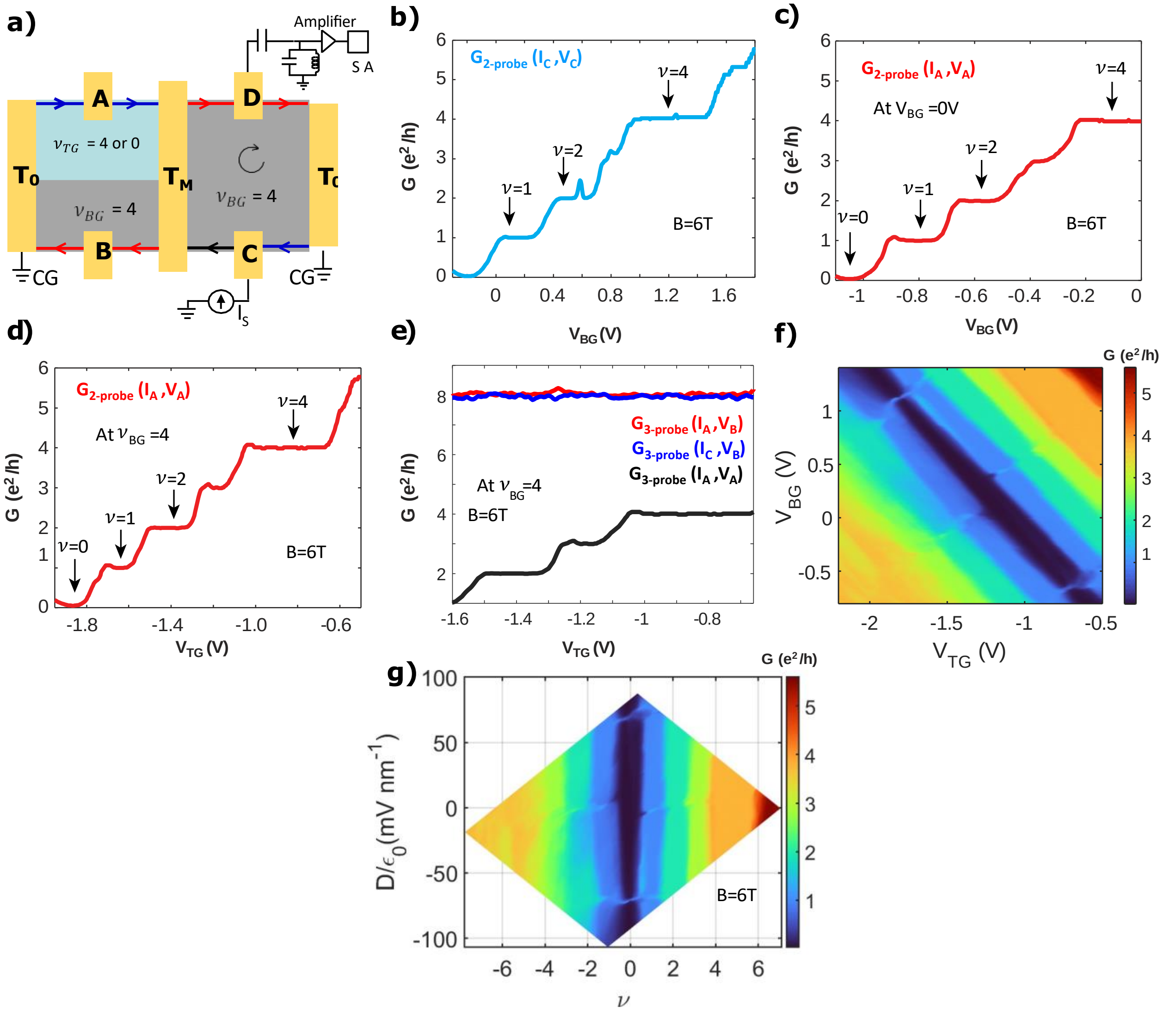

**Supplementary Figure 7: QH response of device 3.** (**a**) Device schematic and measurement setup, which is similar to device 1 and device 2. The only difference here is that the local BLG area (shown in sky blue color) is controlled by the top gate (TG), instead of the local back gate in supplementary Fig. 2 and supplementary Fig. 4. **(b)** QH response at 6T of the global part measured as ($I_C$ , $V_C$ ) with global back gate voltage ($V_{BG}$). **(c)** QH response at 6T of the local part measured as ($I_A$, $V_A$) with local top gate voltage ($V_{TG}$) for $V_{BG}$ = 0V. **(d)** QH response ($I_A$, $V_A$) with $V_{TG}$, while $\nu_{BG}$ = 4. Though the nature of the plateau in Fig. 7d remains similar to Fig. 7c, the position of the plateau in the top gate voltage shifted due to different back gate voltages. **(e)** Red data shows the 3-probe conductance ($I_A$, $V_B$) as a function of $V_{TG}$, measured at contact B when current is injected at contact A while the bulk filling is kept at $\nu_{BG}$ = 4. Similarly, blue data shows for ($I_C$ , $V_B$) with $V_{TG}$ at $\nu_{BG}$ = 4. In comparison, the black line shows 2-probe conductance ($I_A$, $V_A$). The 3-probe transmitted (blue line) and reflected (red line) conductance values are double of the 2-probe conductance (for $\nu$ = 4 plateau) confirming equal partitioning of the current from the floating contact. **(f)** 2D color map of conductance ($I_A$, $V_A$) with $V_{BG}$ and $V_{TG}$ at $B = 6T$. **(g)** 2D color map of conductance ($I_A$, $V_A$) with the displacement field and local filling factor (under the top gate part).

The gap closing of the $\nu = 0$ could be seen around $D^* \sim \pm 0.07 V/nm$, indicating the transition from one kind of insulating phase to another one as reported in literature[1,3–5].

## Section S3: Estimation of displacement fields

For device 1 and device 2 (supplementary Fig. 2b and 4b), it can be seen that the Dirac point or the charge neutrality point under the local back gate part appears at $V_{LG} \sim 0$V. Therefore, the $\nu = 0$ state in device 1 appears at $D \sim 0$ V/nm, and expected to be CAF [6,7]. Due to two graphite back gates (global and local), we could not vary the displacement field, $D$, to access the other phases of the $\nu = 0$ QH state of BLG. In device 3, with the combination of the local top gate and global back gate (shown in supplementary Fig. 6a), we can independently control the charge carrier density and the displacement field in the local BLG part. The applied electric displacement field $D_B$ and $D_T$ resulting from back and top gating is given by $D_B = \frac{\epsilon_b(V_{BG}-V_{BG}^{DP})}{d_b}$ and $D_T = \frac{\epsilon_t(V_{TG}-V_{TG}^{DP})}{d_t}$, where $\epsilon_b$, $\epsilon_t$ are the dielectric constants, $d_b$, $d_t$ are the thickness of the bottom and top hBN (dielectric) layers, and $V_{BG}^{DP}$, $V_{TG}^{DP}$ are the voltages corresponding to the Dirac point in back gate and top gate voltages, respectively. The net displacement field in the bilayer graphene is given by

$$D = \frac{D_B - D_T}{2} = \frac{\frac{\epsilon_b(V_{BG}-V_{BG}^{DP})}{d_b} - \frac{\epsilon_t(V_{TG}-V_{TG}^{DP})}{d_t}}{2} \tag{S1}$$

From supplementary Fig 6b, we note that $V_{TG}^{DP} \sim -1.2$V and $V_{BG}^{DP} \sim 0.25$V. Using the values of $\epsilon_b = \epsilon_t \sim 4$, and $d_t$ = 32nm, and $d_b$ = 50nm (measured using AFM), we have converted the axis of supplementary Fig. 7f (with $V_{BG}$ and $V_{TG}$) into $D$ and $\nu$, and plotted in supplementary Fig. 7g. From supplementary Fig. 7g, one can notice the gap closing of the $\nu = 0$ happens around $D \sim \pm 0.07 V/nm$, which is consistent with the critical $D*$[1, 3 4, 5] required at $B \sim 6T$ to cross from the CAF to FLP[1, 3, 6]. The thermal conductance of the (4,0) state of device 3 was carried out at $D \sim -0.08 V/nm$ ($V_{BG} \sim 1.2V$, $V_{TG} \sim -1.9V$), which was higher than the critical field ($D* \sim -0.07 V/nm$), and thus measured for the FLP phase[3, 6–8]. The corresponding transverse electric field energy or valley anisotropy energy ($E_\nu$) for $D \sim -0.08 V/nm$ is of the order of $\approx 8 - 10 meV$, which is experimentally measured in the literature[3, 9, 10] including our work [5]. The $E_\nu$ can also be theoretically estimated as $E_\nu = \frac{D \times d}{\epsilon_{BLG}}$, where $d$ is the separation between the two layers of BLG which is $\sim 0.35$nm, and $\epsilon_{BLG}$ is the dielectric constant of the BLG. The $\epsilon_{BLG} \sim 4.3$ is consistent with the experimentally measured $E_\nu$[3, 5, 8–10]

## Section S4: Noise measurement setup

For noise measurement, the device was mounted on a chip carrier which was connected to the cold finger fixed to the mixing chamber plate of the dilution refrigerator. Schematic of the noise measurement setup is shown in supplementary Fig. 8, where the sample is set at $\nu$=4. The ground contacts shown in the figure are directly shorted to the cold finger to achieve the cold ground. The sample was current-biased from contact C with a 1 GΩ resistor located at the top of the dilution fridge. Current fluctuations measured at contact D are converted on-chip into voltage fluctuations using the quantum Hall (QH) resistance $R = h/\nu e^2$ where $\nu$ is the filling factor. The voltage noise generated from the device is filtered using a superconducting resonant LC tank circuit with the resonance frequency of ∼725kHz and bandwidth 30kHz. The filtered signal is then amplified by a homemade cryogenic voltage pre-amplifier which was thermalized to 4K plate of dilution refrigerator. This pre-amplified signal was then amplified using a voltage amplifier (PR-E3-SMA S/N 115016) placed at the top of the fridge at room temperature. After the second stage of amplification, the amplified signal was measured using a spectrum analyzer (N9010A). All the noise measurements were done using the bandwidth of ∼ 30 kHz. At zero bias, the equilibrium voltage noise measured at the amplifier contact is given by

$$S_V = g^2(4k_BTR + V_n^2 + i_n^2R^2)BW\,, \tag{S2}$$

where $k_B$ is the Boltzmann constant, $T$ is the temperature, $R$ is the resistance of the QH state, $g$ is the gain of the amplifier chain, and $BW$ is the bandwidth. The first term, $4k_BTR$, corresponds to the thermal noise, and $V_n^2$ and $i_n^2$ are the intrinsic voltage and current noise of the amplifier. At finite bias, the QH edge emanates from the floating contact with a higher quasi-equilibrium temperature. The excess voltage noise $\delta S_V$ is converted to excess current noise $S_I$ according to $S_I = \frac{\delta S_V}{R^2}$, where $R = \frac{h}{\nu e^2}$ is the resistance of the considered QH edge. At the same time, the intrinsic noise of the amplifier remains unchanged. Frequency independence of the thermal noise and the excess noise allows us to operate at higher frequency (∼725 kHz), so that we can eliminate the contribution from flicker noise (1/f) which usually becomes negligible for frequencies above a few tens of Hz. The resonant LC tank circuit was built using inductor L of $\sim$ 365 $\mu$H made from a superconducting coil thermally anchored to the mixing chamber of the dilution refrigerator. The parallel C of $\sim$ 125 pF is the capacitance that develops along the coaxial lines connecting the sample to the cryogenic pre-amplifier. A ceramic capacitance of 10 nF was introduced between the sample and inductor to block the dc current along the measurement line. The typical input voltage noise and current noise of cryogenic pre-amplifiers were $\sim$ 250 pV/$\sqrt{\text{Hz}}$ and $\sim$ 20-25 fA/$\sqrt{\text{Hz}}$ [11, 12], respectively.

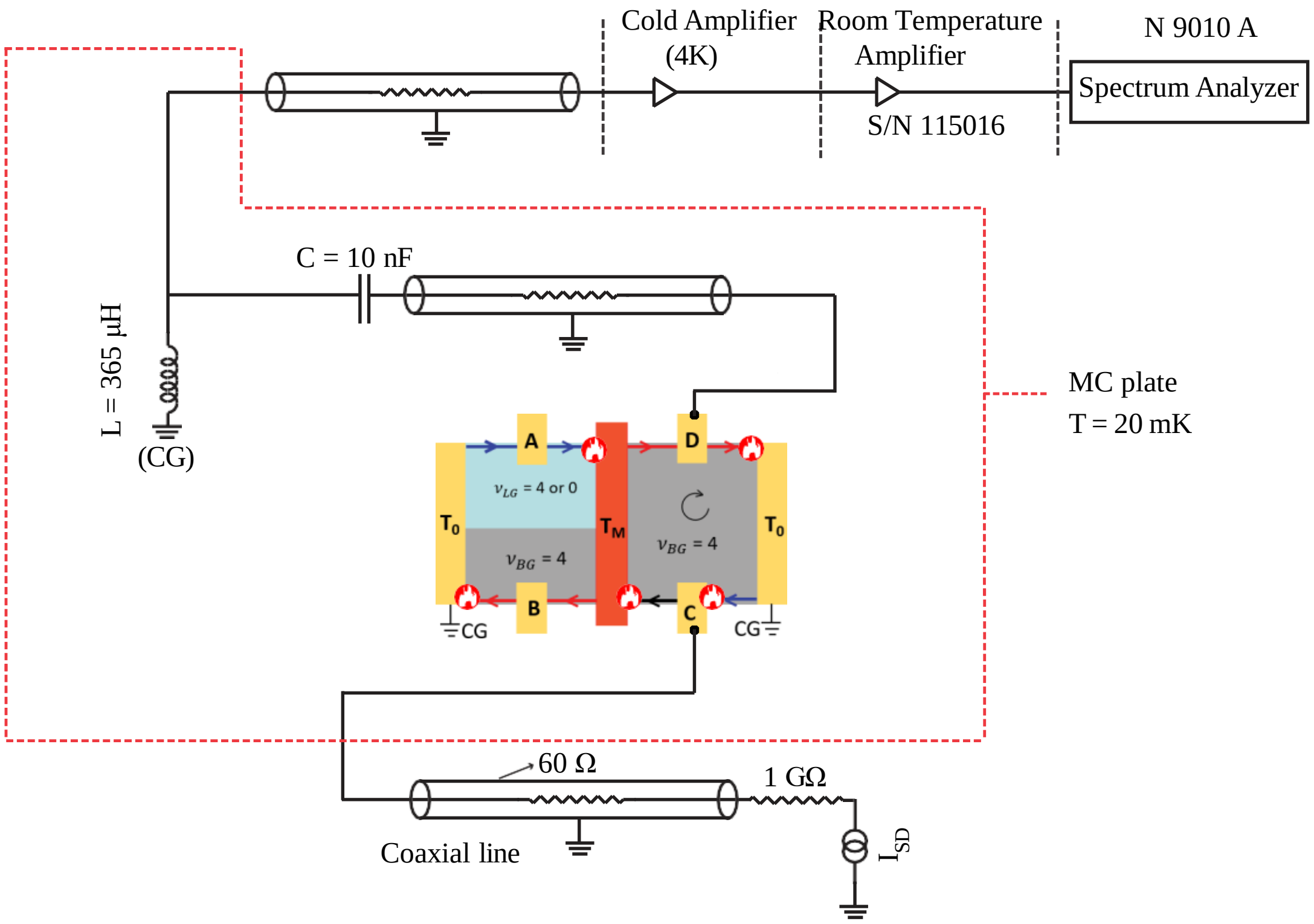

**Supplementary Figure 8:** Experimental setup for noise measurement.

## Section S5: Gain and electron temperature estimation

In our analysis of the noise data, it is crucial to know the gain of the amplifier chain and the electron temperature ($T_e$) of the system. In this section, we describe how we measure these two quantities. Note that at a given bath temperature, $T$ (phonon temperature), the electron temperature, $T_e$ can be different. For our measurement setup, the cooling of the electron system remains efficient down to our lowest bath temperature, and the $T_e$ remains very close to the bath temperature throughout the entire temperature range of our measurements. Details are shown in our previous work[11–13], here we briefly mention how we extract the gain and the $T_e$ corresponds to our base temperature (lowest bath temperature, $20mK$) for the measured devices.

We calculate precisely the gain and $T_e$ by measuring thermal noise at zero current bias. We have estimated the gain of the amplification chain and the electron temperature from temperature-dependent Johnson-Nyquist noise (thermal noise) using equitation S1. At a quantum Hall plateau, any change in bath temperature will only affect the first term in equitation S1, while all other terms are independent of temperature. If one plots the $\frac{S_V}{BW}$ as a function of temperature, the slope of the linear curve will be equal to $4g^2k_BR$. Since at the quantum Hall plateau, the resistance R is exactly known, one can easily extract the gain of the amplification chain from the slope and the intrinsic noise of the amplifier from the intercept. The gain is found using the following equation:

$$g = \sqrt{\frac{\partial \frac{S_V}{BW}}{\partial T} \frac{1}{4k_BR}}\,, \tag{S3}$$

where $\frac{\partial \frac{S_V}{BW}}{\partial T}$ is the slope of the linear fit. The implementation of this procedure is shown for $\nu = 4$ in Supplementary Fig. 9, where, the noise spectrum ($S_V$) at zero impinging current measured at different bath temperatures is shown as a function of frequency (Fig. 9(a)). The $S_V$ value at the resonance frequency, divided by BW, is plotted as a function of bath temperature in Supplementary Fig. 9(b), where the red solid line is the linear fit to the data in the temperature range from 0.1K to 0.8K. From the slope, we extract the gain, which is found to be $\sim$ 1086. Note that we do not use the base temperature data for the fitting in Supplementary Fig. 9(b) because the electron temperature ($T_e$) could be different from the base temperature for such a low temperature.

As the gain is known, one can calculate the $(V_n^2 + i_n^2 R^2)$ from the intercept of the linear fitting of $S_V/BW$ vs temperature. Now from the known value of the measured noise at the base temperature, the corresponding electron temperature ($T_e$) can be found directly using the following equation:

$$T_e = \frac{\frac{S_V}{g^2 BW} - (V_n^2 + i_n^2 R^2)}{4 k_B R}. \tag{S4}$$

The measured value of noise for $\nu = 4$ at the base temperature ($20mK$) corresponds to $T_e = 24mK$, which is consistent with the electron temperature measured in our previous work[12,13]. The fact that $T_e$ is very close to the bath temperature can be also seen directly from Supplementary Fig. 9(b): The $T_{bath} = 20$ mK data point (black open circle) is located almost on the dashed line representing the extrapolation of the linear fit into the region below $0.1$ K.

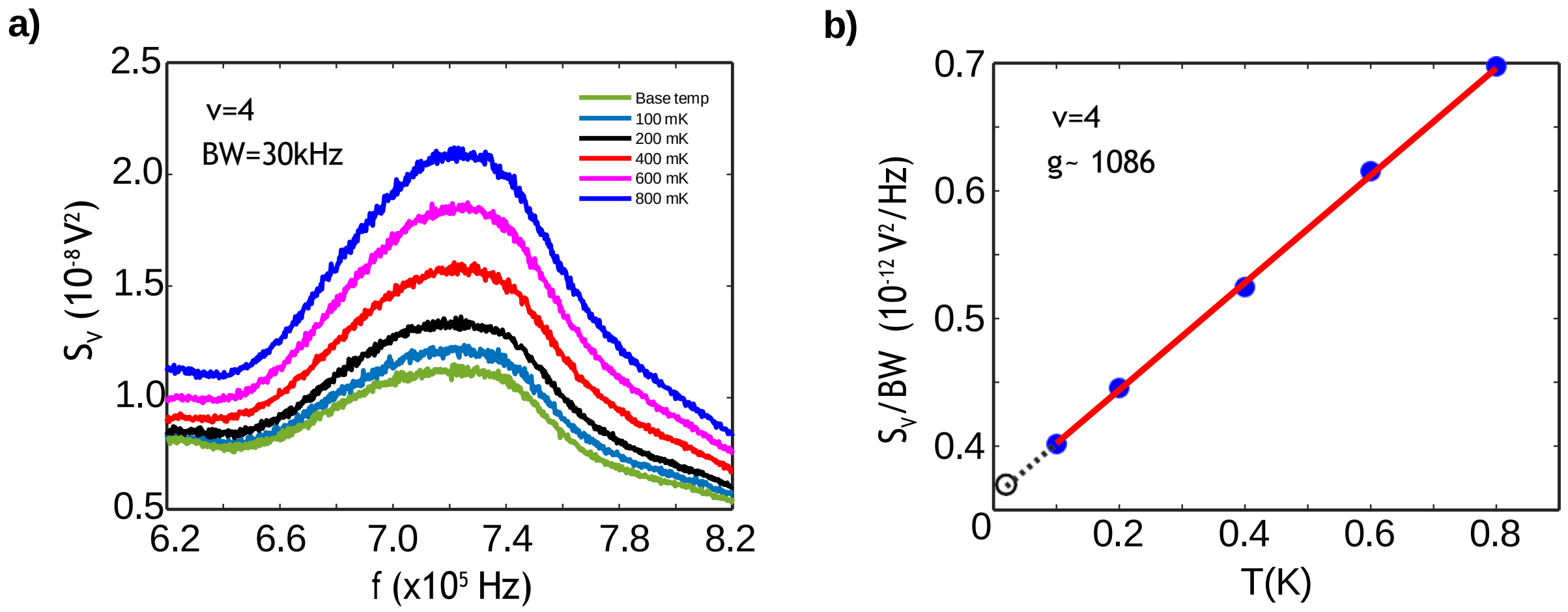

**Supplementary Figure 9: Gain and electron temperature estimation at $\nu = 4$.** (**a**) The voltage noise $S_V$ (see equation S2) measured by a spectrum analyzer is plotted as a function of the frequency at different bath temperatures for $\nu = 4$. From this plot, the resonance frequency of the tank circuit was found to be $\sim$725 kHz. (**b**) Blue solid circles represent the noise $S_V$ divided by bandwidth ($BW$) at the resonance frequency as a function of the bath temperature from 0.1K to 0.8K. The open black circle represents the data for the base temperature ($20mK$). The solid red line is a linear fit to the data from $0.1$ K to $0.8$ K, and the dashed line is the linear extrapolation below $0.1$ K. From the slope of the linear fit (using equation S3), the gain $g$ was found to be $\sim$1086.

**Section S6: Phase diagram of $\nu = 0$ state of BLG:**

As described in the references [6, 7], the $\nu = 0$ state of BLG is theoretically believed to host four distinct ground states, each characterized by different spin and valley (equivalent to layer or sublattice) degrees of freedom. These ground states are ferromagnetic (F), canted antiferromagnetic (CAF), partially layer polarized (PLP), and fully layer polarized (FLP). While all four phases are electrically insulating in bulk, the ferromagnetic phase is expected to exhibit conducting helical edge modes at the physical boundary of a device. According to the Ref. [7], the ground state of the $\nu = 0$ phase in BLG is determined by competing interaction energies: $u_z$, $u_\perp$, $E_z$, and $E_v$. Here, $u_z$ and $u_\perp$ represent the isospin anisotropy energies in the z-axis and the x-y plane, respectively, and are known as Kharitonov parameters. $E_z$ is the Zeeman energy, which is proportional to the total magnetic field $B_{tot}$, and $E_v$ is the valley anisotropy energy, which is proportional to the applied displacement field ($D$). These four phases are separated by phase boundaries and converge at a single quadruple point. The position of quadruple point ($u_\perp = -E_z/2$, $u_z = E_v - E_z/2$) depends on the values of $E_z$ and $E_v$.

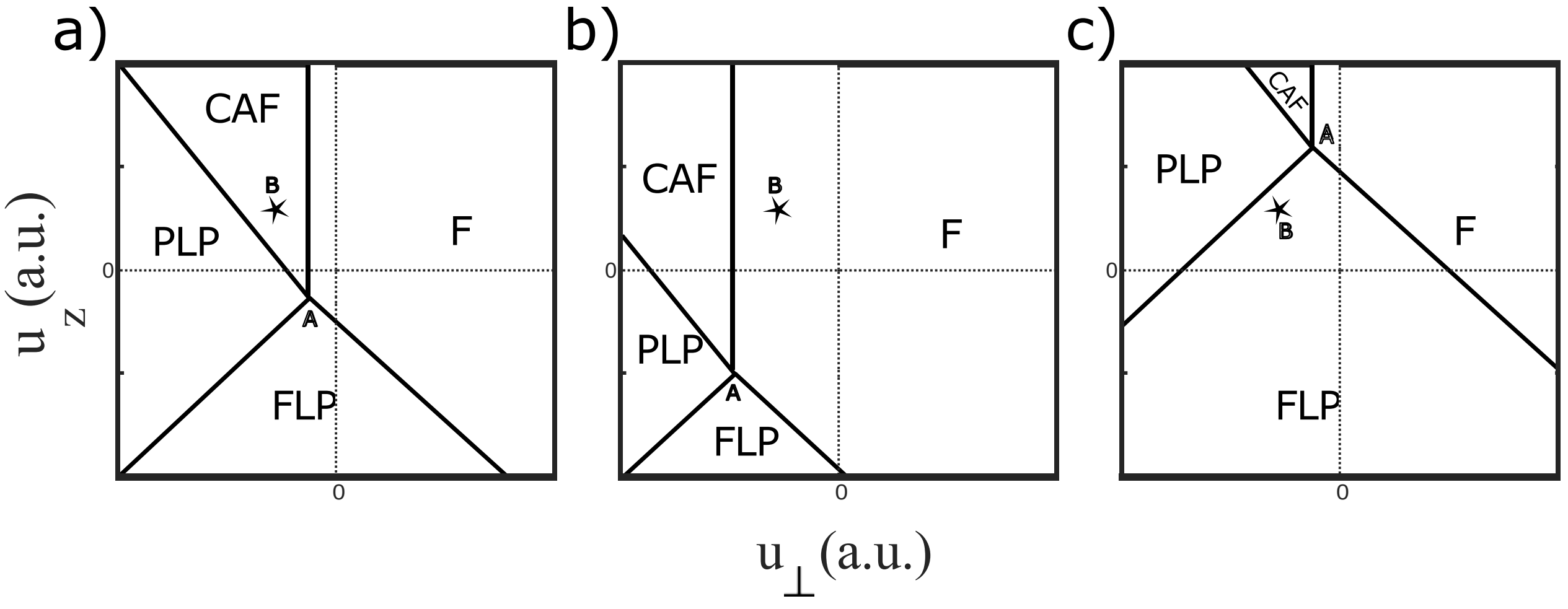

**Supplementary Figure 10: Phase diagram of $\nu = 0$ state of BLG.** (**a**) The schematic representation of different phases as a function of isospin anisotropy energies, $u_z$ and $u_\perp$ [7]. The solid lines indicate the phase boundaries or phase transition lines. The point '**A**' represents the quadrupole point, where four phases converge to a single point. The point '**B**' marked as a star represents the position of a sample in the CAF phase of the phase diagram for a given $u^*_z$ and $u^*_\perp$. (**b**) The schematic representation of the phase diagram in the presence of the large in-plane magnetic field (which shifts the quadrupole point, **A** along the diagonal direction) such that the sample lies in **F** phase. (**c**) The schematic representation of the phase diagram in the presence of the large displacement field (which shifts the quadrupole point, **A** along the vertical direction) such that the sample lies in **FLP** phase.

To determine the ground state (or the Kharitonov parameters) of the $\nu = 0$ state in BLG, it is necessary to cross the phase boundaries, resulting in a phase transition. Since $u_z$, $u_\perp$, and $E_z$ all scale linearly with the perpendicular magnetic field ($B_\perp$), changing its value does not induce any phase transition. This makes the experimental determination of $u_z$ and $u_\perp$ challenging. However, by varying the in-plane magnetic field (which affects only the Zeeman energy) or the displacement field, one can shift the position of the quadruple point while keeping the $u_z^*$ and $u_\perp^*$ of a device fixed, as shown schematically in supplementary Fig. 10. This induces transitions between the phases, as summarized by Pientka [6]. As illustrated in supplementary Fig. 10b, a large in-plane magnetic field (large Zeeman energy) can realize the ferromagnetic (F) phase, which is conducting and thus easy to identify without ambiguity. From the critical in-plane magnetic field ($B_{tot}^*$) required, one can estimate the value of $u_\perp$. For typical devices with a similar dielectric environment (both sides hBN encapsulated with a thickness of 30-40 nm and graphite-gated devices), the reported $B_{tot}^*$ for BLG devices is around 15-20 T [1, 3], while the perpendicular magnetic field ($B_\perp$) is around 3-5 T. This corresponds to $u_\perp^* \approx -0.6$ to $-0.7$ meV, which is similar in magnitude to that for single-layer graphene (SLG) devices [14, 15]. Determining the value of $u_z^*$ is quite challenging. As shown schematically in supplementary Fig. 10c, by varying $E_v$, a transition to the fully layer polarized (FLP) phase (which is insulating) can occur at a critical $E_v^*$. By following the phase boundary from the ferromagnetic (F) phase to the FLP phase, one can estimate $u_z^*$. According to literature [1, 3, 8], the typical values of $u_z^*$ for BLG devices range from approximately 1.6 to 4.5 meV, which is similar to values reported for single-layer graphene (SLG) devices [14, 15].

According to the phase diagram in Ref. [7], these typical values ($u_z^*$ and $u_\perp^*$) suggest that the canted antiferromagnetic (CAF) phase is expected to be the ground state. However, the alignment of hexagonal boron nitride (hBN) with SLG or BLG samples can alter the nature of the ground state. This alignment induces a sub-lattice potential (valley Zeeman coupling), shifting the quadruple point vertically in supplementary Fig. 10c, similar to the effect of an external electric field ($E_v$) in BLG. Consequently, the ground state could become either the partially layer polarized (PLP) or FLP phase, depending on the degree of alignment [15–17]. Since our measured devices have a similar dielectric environment (both sides encapsulated with hBN of similar thickness, 30-50 nm, and graphite-gated) as reported in the literature, and for device 1 and device 2, the hBN layers are not aligned (as indicated by the temperature-dependent resistance data shown in supplementary Fig. 3), we expect $u_\perp^*$ and $u_z^*$ for our BLG devices to be in the range of approximately -0.6 to -0.7 meV and 1.6 to 4.5 meV, respectively, for the applied magnetic field range of 4-6 T ($B_\perp = B_{tot}$). These values suggest that, theoretically, the canted antiferromagnetic (CAF) phase is expected to be the ground state for our devices.

However, modifications to the Kharitonov phase diagram have been theoretically studied using the Hartree-Fock approximation for SLG [18], revealing the possibility of five phases instead of four, with a new phase exhibiting a coexistence of CAF and bond order, BO (which is similar to PLP for BLG). Further studies [19] have shown that relaxing the stringent ultra-short-range interaction constraints could result in many ground states for $\nu = 0$ in SLG. Therefore, although the CAF phase is expected for our devices (device 1 and device 2), variations in parameters and considerations beyond ultra-short-range interactions may result in the ground state being in nearby phases such as PLP or a coexistence phase like CAF+PLP.

Several experimental attempts have been made to trace ground states, whether CAF or PLP, but the insulating nature of these phases makes this very challenging. The most convincing experiments suggesting that the CAF phase is the ground state involve magnon transport through the $\nu = 0$ state, as reported in Ref.[20, 21] for SLG and in Ref.[8] for BLG. However, these results do not rule out the possibility of a coexis-tence phase of CAF+BO or CAF+PLP as the ground state in SLG and BLG, respectively. Recently, STM experiments [17, 22, 23] have confirmed the BO (like PLP in BLG) as the ground state in SLG. However, since STM (non-magnetic tip) cannot detect the magnetic moment, it is also possible that the observed phase could be CAF +BO. It is important to note that in STM studies, the samples are encapsulated only by bottom hBN, which may create a different dielectric environment than transport studies where SLG or BLG samples are encapsulated by both top and bottom hBN.

## Section S7: Expected thermal conductance response from different ground states:

In our experiment, we aimed to detect these phases via thermal conductance measurements. For devices 1 and 2 at $E_\nu \approx 0$, and for device 3 at larger $E_\nu$, we observed the absence of thermal conductance for the $\nu = 0$ state of BLG. Specifically, for device 3 at $E_\nu \approx 10\text{meV}(D \approx -80\text{mV/nm})$, the expected ground state is FLP [3, 6–8]. The excitation spectrum of this phase is gapped ( 1-2K as per [6]), which explains the observed zero thermal conductance for device 3. For devices 1 and 2, at $E_\nu \approx 0$ and not aligned with hBN, the absence of detectable thermal conductance suggests the following scenarios:

(i) The ground states of devices 1 and 2 are unlikely to be F phase since the measured states are electrically insulating. The phase cannot also be FLP since the BLG devices are not aligned with hBN, and measurements are performed at zero displacement field [6, 7].

(ii) The possible ground states could be CAF or PLP or a coexistence CAF+PLP phase[18]. This is plausible as slight variations in $u_z$ and $u_\perp$ (Kharitonov parameters) could lead the devices into any of these phases, which share a common phase boundary or these phases are located in close proximity [18].

(iii) Although Ref.[6] indicates that collective excitations for the PLP phase are gapless and thus expected to show finite thermal conductance. However. it is also noted that these calculations do not consider higher-order interactions that break U(1) isospin symmetry down to C3 symmetry, which is broken at the lattice scale. Thus, it may no longer remain a continuous symmetry-breaking phase to exhibit gapless Goldstone modes [18, 24, 25]. The exact magnitude of the gap of the excitation spectrum is not well-known in the literature. A gapped spectrum is also expected for the coexistence CAF+PLP phase [18]. Thus, these gaps can hinder the heat flow of the PLP phase (or CAF+PLP) at low temperatures.

(iv) For the CAF phase, gapless excitations are anticipated due to spontaneously broken continuous symmetry, resulting in finite thermal conductance. However, the spectrum of CAF may be gapped due to the quantization of collective modes arising from the finite size of the device, an effect observed in BLG devices [8]. In the following section, we estimate the expected excitation gap for our geometry, which is a fraction of a Kelvin. These gaps will significantly suppress the thermal conductance contribution at low temperatures, hindering the heat flow of a CAF phase.

In summary, at low temperatures, the gaps in the PLP (due to C3 symmetry breaking) or CAF (due to quantization from finite size) phases inhibit heat transport, resulting in the absence of thermal conductance. At elevated temperatures, these gaps can be overcome. However, at higher temperatures, phonons become the dominant heat carriers, as shown in Fig. 5 of the manuscript. At 1K, the phonon contribution to thermal conductance is an order of magnitude higher than the expected contribution from collective excitations [6]. Thus, the thermal conductance contribution from the collective excitations of $\nu = 0$ state could not be measured. This highlights the need for further theoretical and experimental studies to fully understand the nature of excitations in the $\nu = 0$ state and their impact on thermal transport.

### Section S8: Estimations of the gap of collective excitations

Before we discuss how the collective excitations of CAF can be gapped, let us compare the values of thermal conductance from the electronic channel versus its magnitude for the gapless excitations. The thermal conductance for $N = 1$ QH edge channel[11] is given by $G_Q^e \sim 1pW/K^2 \times T$, where $T$ is the temperature. At $100mK$, $G^e_Q \sim 0.1pW/K$. The magnitude of the gapless excitation[6] is given by $G_Q^{CAF} \sim WT^2/v$, where $W$ and $v$ represent the width of the device and the velocity of the excitations, respectively. For $W \sim 5\mu m$, $G_Q^{CAF} \sim 0.17pW/K$ at $100mK$ as estimated in Ref[6]. Thus, at $T \sim 100mK$, the electronic contribution and the contribution from the collective excitations are in a similar magnitude[6]. Therefore, as discussed in the previous section, the inability to observe the measurable heat conductance from the most expected CAF phase might result from the gapped spectrum of the collective excitations stemming from collective mode quantization due to the finite size of the device[8]. The effect of such quantization has already been observed in BLG devices[8]. Following a similar approach, we can also estimate the excitation gap due to finite-size quantization. Considering the model of a one-dimensional Fabry-Perot (FP) cavity, the typical mode energy is given by $E_1 = \frac{\hbar v_{CAF}\pi}{W}$ (Ref. 8), where $v_{CAF}$ is the velocity of the CAF mode and $W$ is the width of the FP cavity. In our devices, the region of $\nu = 0$ is minimally bounded by a length scale of $2 - 4\ \mu m$. Using these length scale parameters and the typical $v_{CAF} \approx 57\ km/s$ (Ref. 8), the excitation gap due to confinement was found to be an order of $350 - 680$ mK, respectively. Thus, these gaps will significantly suppress the thermal conductance contribution at low temperatures, hence hindering the heat flow in the $\nu = 0$ CAF phase.

**Section S9: Thermal interface resistance:**

The absence of thermal conductance in the $\nu = 0$ state observed in our experiment could potentially be due to thermal interface resistance between the hot electron bath of the floating metal and collective excitations (Fig. 1b of the manuscript). However, we argue below that this is unlikely in our setup:

(i) The hot reservoir or hot electron-hole excitations from the floating contact can couple to the localized electrons at the sub-lattice points of the BLG via hopping at the interface (supplementary Fig. 11a). This allows energy exchange, and while hopping from metal to sub-lattice points, spin-flip processes can excite spin wave like collective excitations. The reverse spin-flip process will help to absorb these excitations at the interface. At a steady state, the total angular momentum change of the metal and BLG will be zero, but these dynamics are allowed as long as hopping is possible. Further, the Coulomb interaction of the $\nu = 0$ state of graphene may play a role in energy exchange (spin flip) between the hot electron bath and the collective excitations. Moreover, due to the electron-hole type excitonic nature of the collective excitations, as established in the literature[26], the coupling between the hot electron bath and the collective excitations in BLG is expected. In fact, at lower temperatures, the coupling between the hot electron bath and the collective excitations is expected to be stronger than between the hot electron bath and phonons.

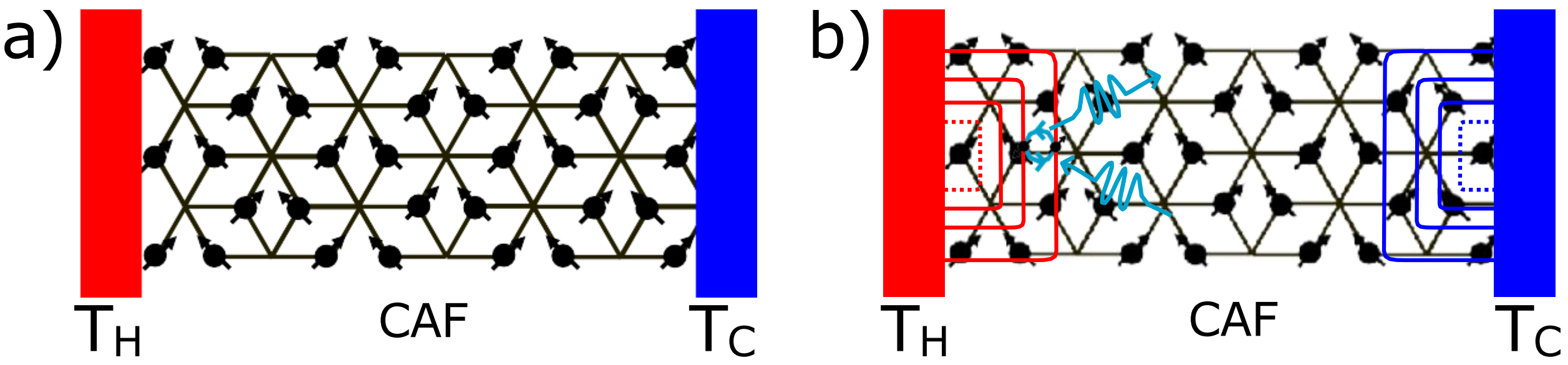

**Supplementary Figure 11: Schematic for coupling between the hot metal and collective excitations.** (**a**) A bilayer graphene is connected by the hot and cold metallic reservoirs. The localized electrons of the $\nu = 0$ state of the BLG are presented by solid black circles with spin, which are located at the sub-lattice points (shown for CAF). The localized electrons near the interface can hop between the metal and BLG. (**b**) The schematic presentation for a BLG, which is locally n-doped, having confined edge states near the contacts. The electron hopping between edge states with opposite spin can couple it to the collective excitations shown by wiggly lines, where the direction of the arrow shows whether it is generated or absorbed.

The transparent coupling between the metal and graphene is evident in our experiment from the resistance data at zero magnetic field and QH response showing negligible contact resistance, as well as from the thermal conductance measurement at integer fillings. Details can be found in our earlier work (Table S3 in the supplementary information of Ref.[11]), which shows the interface between the metal contacts and graphene is more than 99% transparent. This is expected because the one-dimensional graphene edge contacts between the metal and carbon atoms [27] make high-quality electrical contact without a tunnel barrier. Although this is experimentally established for the conducting state of graphene, it is known that the one-dimensional graphene edge makes good coupling between the dangling carbon atoms and the metal at the interface, enabling the hopping of electrons between the metal and BLG.

(ii) Additionally, the metallic contacts used in our experiments tend to locally n-dope the graphene flake, as established by the ability to generate and detect magnons[20, 26, 28]. Consequently, the reservoir of hot electrons extends into the graphene flake itself, potentially enhancing the coupling to collective modes. This is shown schematically in supplementary information Fig. 11b, where local doping confines the edge states near the contacts, and the tunneling of electrons between the hot electronic edge states with opposite spins can generate and absorb collective excitations via spin-flip processes. This scenario likely enhances the coupling between the hot reservoir and the collective modes in a real device.

(iii) Our experimental design to study the bulk heat flow is motivated by the Ref.[6], where the bulk thermal conductance of CAF/PLP phase was theoretically proposed using hot and cold metallic leads connected to a bilayer graphene flake, imposing their respective temperatures on the collective modes on either side of the sample. No thermal interface resistance between the hot metal and collective excitations is considered in Ref. [6].